\documentclass[aip,jcp,preprint,numerical,floatfix]{revtex4-1}

\usepackage{amsmath,amssymb,amsfonts}
\usepackage{mathrsfs}
\usepackage{epsfig}
\usepackage{bm}
\usepackage{rotating}
\usepackage{graphicx}
\usepackage{eucal}
\usepackage{appendix}
\usepackage{footnote}
\usepackage{xcolor}
\usepackage[flushleft]{threeparttable}

\begin{document}



\title{Eckart ro-vibrational Hamiltonians via the gateway Hamilton operator: theory and practice}

\author{Viktor Szalay}

\email[Email address:~]{szalay.viktor@wigner.mta.hu}

\affiliation{Institute for Solid
State Physics and Optics, Wigner Research Centre for Physics, Hungarian Academy of Sciences, P. O.
Box 49, H-1525 Budapest, Hungary }

\date{\today}

\begin{abstract}
Recently, a general expression for Eckart-frame 
Hamilton operators has been obtained by the gateway Hamiltonian method ({\it J. Chem. Phys.} {\bf 142}, 174107 (2015); 
{\it ibid.} {\bf 143}, 064104 (2015)). The kinetic energy operator 
in this general Hamiltonian is nearly identical with that of the Eckart-Watson operator even when 
curvilinear vibrational coordinates are employed. Its different realizations 
correspond to different methods of calculating Eckart displacements.
There are at least two different methods for calculating such displacements: rotation and projection. In this communication 
the application of Eckart Hamiltonian operators constructed by rotation and projection, respectively, is 
numerically demonstrated in calculating vibrational energy levels. 
The numerical examples confirm that there is no need for rotation to construct an Eckart ro-vibrational Hamiltonian. 
The application of the gateway method is advantageous even when rotation is used, since it obviates the need for 
differentiation of the matrix rotating into the Eckart frame.
Simple geometrical arguments explain that there are infinitely many different methods for calculating Eckart displacements. 
The geometrical picture also suggests that a unique Eckart displacement vector may be defined 
as the shortest (mass-weighted) Eckart displacement vector among Eckart displacement vectors
corresponding to configurations related by rotation. Its length, as shown analytically and demonstrated by way of numerical 
examples,  is equal to or less than 
that of the Eckart displacement vector one can obtain by rotation to the Eckart frame.

\end{abstract}

\maketitle

\section{Introduction} \label{introduction}

The importance of  Eckart conditions
and Eckart ro-vibrational Hamiltonians  cannot be overestimated.   
Their importance is described succinctly by Sutcliffe\cite{sutcliffe-hinchliffe}
\begin{description}
\item "It would, I believe, be widely agreed that the modern theory of molecular spectra 
began with publication by Carl Eckart in 1935 of his paper 
{\sl Some Studies Concerning Rotating Axes and Polyatomic Molecules}\cite{eckart}. 
It would be probably also be widely agreed that the apogee of this work occurred 
in 1968 when James K. G. Watson\cite{watson}, published {\sl Simplification of the molecular vibration-rotation hamiltonian} 
which put Eckart's classical mechanical form into a proper quantum mechanical one. 
This leads to the wave mechanical problem for molecular vibration-rotational motion specified by what 
we shall call the Eckart-Watson Hamiltonian."
\end{description}
Nearly at the same time of Watson's work, the exact, general form of 
the quantum mechanical rotational-vibrational kinetic energy operator (KEO) given in curvilinear vibrational 
coordinates has been established by Meyer and G\"unthard\cite{meyer-gunthard}, and Pickett\cite{pickett}. 
However, accounting for Eckart's rotating axes in this general KEO has turned out to be difficult 
and has remained the subject of ongoing research
\cite{wei-carrington-1,wei-carrington-2,mardis-sibert,wei-carrington-3,rey,wei,meremianin-1,pesonen,
mccoy-burleigh-sibert,wang-sibert,szidarovszky-fabri-csaszar,
wang-carrington,fabri-matyus-csaszar,sadri,yachmenev-yurchenko,lauvergnat-luis}.
Research on how to employ the general ro-vibrational KEO possibly by accounting 
for the Eckart conditions has gained practical motivation with the
advance of experimental spectroscopic studies of highly excited vibrational sates  
and large-amplitude internal motions of molecules \cite{h5+,ch5+,ch4h2o-1,ch4h2o-2}.        

If it is so difficult to use, one may wonder, why to use Eckart conditions at all. 
One reason is that they minimize Coriolis coupling, thereby they give justification of thinking 
about the nuclear motions of a molecule as rotational and vibrational motions. With the Coriolis terms 
minimized one might also expect that fewer basis functions may suffice to obtain converged 
ro-vibrational energy levels in a variational calculation\cite{wang-carrington,yachmenev-yurchenko}. 
The use of Eckart frame 
may be the best choice in evaluating dipole-moment matrix elements\cite{sueur}. 
Furthermore, applications of the Eckart frame are not restricted to small and medium size molecules, but it has
been employed in describing nuclear motion dynamics of biomolecules, e.g. proteins \cite{noguti-gon}.

The current situation on using Eckart frame is well described by Wang and Carrington\cite{wang-carrington}
\begin{description}
\item "..., it appears that there is a massive gulf between knowing that it would be best to use an Eckart frame 
and developing a theoretical/computational scheme for exploiting the Eckart 
advantage while at the same time using curvilinear coordinates that enable one 
to cope with large amplitude motion. When normal coordinates are used it is 
straightforward to use an Eckart frame, however, the Eckart KEO in internal coordinates 
is very complicated. Such KEOs have been derived for triatomic molecules, but never employed to compute spectra. 
This conundrum is resolved by computing G-matrix elements numerically."
\end{description}
Although there have been  interesting theoretical developments, 
such as the application of geometric algebra to derive Eckart KEOs \cite{pesonen}, 
and a method of analytical differentiation of the rotation matrix transforming 
into the Eckart frame has been introduced\cite{wang-song,changala,changala-baraban}, practical application of the Eckart frame 
has remained brute force numerical work considerably complicated by employing an energy operator with
little resemblance of the Eckart-Watson Hamiltonian. 
All approaches to incorporate the Eckart conditions into the nuclear motion Hamiltonian 
have assumed explicitly\cite{wang-carrington,fabri-matyus-csaszar,sadri,yachmenev-yurchenko,lauvergnat-luis} 
or implicitly\cite{pesonen} 
a rotation matrix determined such that the Eckart-axis conditions\cite{strauss,redding,meyer,redding2,kudin,
dierksen,dymarsky,galbraith,krasnos}, 
which should not be mistaken for the (rotational) Eckart conditions, be satisfied. 
However, studies on the gateway Hamiltonian method\cite{szalay-eckart,szalay-gateway,szalay-aspects} 
have shown another solution to this question: Projection.
   
The purpose of the present work is to simplify and advance the use of Eckart conditions in variational calculations
of ro-vibrational energy levels by employing the gateway Hamiltonian method. 

The KEO in the general Eckart Hamilton operator 
as obtained by the gateway method is of an expression nearly identical 
with that of the Eckart-Watson operator even when curvilinear vibrational coordinates are employed. 
Its use, as will be numerically demonstrated, 
\begin{itemize}
\item obviates the need for differentiation of the matrix rotating into the Eckart frame,
\item allows construction of an Eckart ro-vibrational KEO without rotation,
\item and leads to defining optimal Eckart displacements.
\end{itemize} 

The outline of the paper is as follows. In Section II the basic ideas and formulas of 
the gateway Hamiltonian method  are reviewed. A method of solving 
the vibrational Schr\"odinger equation corresponding to the general Eckart Hamilton operator is described in Section III. 
Numerical examples are given in Section IV to demonstrate that: 
a) There is no need for differentiating the matrix rotating into the Eckart frame. 
b) The rotation and projection methods of calculating Eckart displacements lead to different Hamiltonians. 
c) Although they are different, these Hamilton operators  have the same spectrum. 
By considering the geometry associated with  the Eckart conditions it is explained  
why there are infinitely many different Eckart displacements (and Eckart Hamiltonians) and 
how optimal Eckart displacements may be defined (Section V). Numerical examples of calculating optimal 
Eckart displacements are also given. Section VI summarizes the results. Two important questions, 
but mainly mathematical in character, are detailed in Appendices. 
In Appendix \ref{eckart-sol}  analytical solution of the Eckart conditions is presented, 
while Appendix \ref{iterativemin} describes the method employed to calculate optimal Eckart displacements. 

\section{The general-form Eckart ro-vibrational Hamilton operator} \label{gateway-h}

The derivation of  the nuclear motion Hamilton operator of a molecule of $N$ atoms starts by relating 
the Cartesian laboratory system (LS) coordinates,
$u_{\overline{\alpha}n}$  $\left( \overline{\alpha}=X,Y,Z;~~n=1,2,\ldots,N \right)$, 
of the atoms of the molecule to the  
translational,
$R_{\overline{\alpha}}$,
rotational, 
$\theta,\phi,\chi$,
and  vibrational,
$s_{r},r=1,2,\ldots,3N-6$, coordinates of the molecule (assumed to be non-linear) by the equation
\begin{gather}
u_{\overline{\alpha}n} =
R_{\overline{\alpha}}+ \left[ \bm{S}^{-1}\left( \theta,\phi,\chi\right)\right]_{\overline{\alpha}\beta}
a_{\beta n}\left( \left\{ s_{r} \right\} \right), \label{rovibcoords}
\end{gather}
where $\beta = x,y,z$ label the axes of a molecule fixed Cartesian system (MS), $\bm{S}\left( \theta,\phi,\chi\right)$ 
is a real rotation matrix parametrized by Euler angles, 
\begin{gather}
a_{\beta n}\left( \left\{ s_{r} \right\} \right)=a_{\beta n}^{0}
+d_{\beta n} \left( \left\{ s_{r} \right\} \right), \label{specdisp}
\end{gather}
\begin{gather}
\sum_{n} m_{n} a_{\beta n}^{0} =0, \label{refconfcent}
\end{gather}
and the displacements $d_{\beta n} \left( \left\{ s_{r} \right\} \right)$ are as yet unknown functions of the
coordinates $s_{r}$. Summation over repeated Greek indices is assumed 
here and throughout the paper. 
$a_{\beta n}^{0}$ are the coordinates of the atoms of a reference configuration.
$m_{n}$ denotes the mass of the $n$th atom of the molecule.
The origin of the MS is fixed to the center
of mass of the reference configuration and its
axes can be oriented and fixed in any convenient way to the reference configuration. Thus the MS has been completely determined and 
the atom coordinates, $a_{\beta n}\left( \left\{ s_{r} \right\} \right)$, 
of any distorted configuration must be given with respect to this MS.

{\sl We aim at determining the $3N$ unknown functions $d_{\beta n}\left( \left\{ s_{r} \right\} \right)$ 
	( or $a_{\beta n}\left( \left\{ s_{r} \right\} \right)$ ).}

This goal will be achieved in two steps: 1) A general form of displacements obeying the Eckart  conditions will be established. 
2) Then it is shown how dependence of these displacements on vibrational coordinates can be introduced.  

\subsection{The Eckart conditions} \label{eckartcond}
  
Together with the coordinates of overall translation and rotation there are $3N+6$ unknowns, but there are only
$3N$ equations in Eq.\, (\ref{rovibcoords}). Since one must be able to calculate the coordinates
of nuclear motion 
from given LS coordinates
and vica versa,
we must complete Eq.\, (\ref{rovibcoords}) with six additional equations. 
Three equations are obtained immediately by noting that Eqs.\, (\ref{rovibcoords}) 
imply that $a_{\beta n}\left( \left\{ s_{r} \right\} \right)$
are atom coordinates in a translation reduced configuration space, that is
\begin{gather}
\sum_{n}m_{n}a_{\beta n}\left( \left\{ s_{r} \right\} \right)=0
\end{gather}
must hold. Then by considering Eqs.\ (\ref{specdisp}) and (\ref{refconfcent}) one finds that the displacements must obey three 
equations
\begin{gather}
\sum_{n}m_{n}d_{\beta n}\left( \left\{ s_{r} \right\} \right) =0 \label{teckart}.
\end{gather}
These equation are called translational Eckart conditions. Three conditions are still missing. They  
determine the functions $a_{\beta n}\left( \left\{ s_{r} \right\} \right)$ 
(and $d_{\beta n}\left( \left\{ s_{r} \right\} \right)$  ),
that is the coordinates of distorted configurations in the MS. 

The simplest conditions are geometrical and somewhat ad hoc.
For instance, having fixed a right-handed MS to the reference configuration 
such that the $x$ axis is parallel to the bond connecting the atoms $A$ and $B$ 
and the $y$ axis lies in  
the plane containing the atoms $A,~B$, and $C$ and points to the half-plane containing $C$, 
we may require these conditions to hold even for distorted configurations.    

To obtain the missing equations in more reasonable way let us consider the expression of the
classical kinetic energy \cite{szalay-eckart}:
\begin{widetext}
\begin{eqnarray}
T &=& \frac{1}{2} M \frac{d\bm{R}}{dt}\cdot 
			\frac{d\bm{R}}{dt}  + \left( \frac{d\bm{S}}{dt} \frac{d\bm{R}}{dt}\right) \cdot
		\left[ \sum_{n=1}^{N} m_{n} \left( \bm{a}_{n}^{0}+ \bm{d}_{n}\right)\right] 
		\nonumber  \\
&+& \left( \bm{S}\frac{d\bm{R}}{dt}\right)\cdot \frac{d}{dt}
		\left[ 
		\sum_{n=1}^{N} m_{n} \left( \bm{a}_{n}^{0}+ \bm{d}_{n}\right) \right] 
		\nonumber \\
&+& \frac{1}{2}\bm{\omega}\cdot \bm{\omega} 
			\sum_{n=1}^{N} m_{n}\left( \bm{a}_{n}^{0}+ \bm{d}_{n}\right)
			\cdot \left( \bm{a}_{n}^{0}+ \bm{d}_{n}\right)-\frac{1}{2} 
			\sum_{n=1}^{N} m_{n}\left( \bm{\omega} \cdot
			\left( \bm{a}_{n}^{0}+ \bm{d}_{n}\right) \right)^{2}   \nonumber \\
&+&  \bm{\omega} \cdot
			\sum_{n=1}^{N} m_{n} \left(\bm{d}_{n}\times \frac{d \bm{d}_{n}}{dt}\right) 
		+\bm{\omega} \cdot \frac{d}{dt}\left[ \sum_{n=1}^{N} m_{n} \left( \bm{a}_{n}^{0} 
			\times \bm{d}_{n}\right) \right]  \nonumber \\
&+& \frac{1}{2}
			\sum_{n=1}^{N} m_{n}\frac{d \bm{d}_{n}}{dt}\cdot 
			\frac{d \bm{d}_{n}}{dt},  \label{classick}
\end{eqnarray}
\end{widetext}
where $\bm{\omega}$ is 
an angular velocity vector whose components are defined as elements of the skew symmetric
matrix ${\bm{S}}\frac{d \bm{S}^{-1}}{dt}$. 
$\bm{d}_{n}$ is defined as a column vector, $\bm{d}_{n} = \left( d_{xn},d_{yn},d_{zn}\right)^{T}$, 
with superscript $T$ denoting transposition. 
Its dependence on the vibrational
coordinates is, for simplicity, not indicated explicitly. Finally, 
$\bm{a}_{n}^{0} = \left( a_{xn}^{0},a_{yn}^{0},a_{zn}^{0}\right)^{T}$. The $ \cdot$ means dot product.
Note that by requiring satisfaction of the translational Eckart conditions and the three equations 
   \begin{gather}
   \sum_{n=1}^{N} m_{n} \bm{a}_{n}^{0} \times \bm{d}_{n} =0 \label{reckart}
   \end{gather}
the expression of kinetic energy simplifies to
\begin{widetext}
\begin{eqnarray}
T &=& \frac{1}{2} M \frac{d\bm{R}}{dt}\cdot 
\frac{d\bm{R}}{dt}   
\nonumber \\
&+& \frac{1}{2}\bm{\omega}\cdot \bm{\omega} 
\sum_{n=1}^{N} m_{n}\left( \bm{a}_{n}^{0}+ \bm{d}_{n}\right)
\cdot \left( \bm{a}_{n}^{0}+ \bm{d}_{n}\right)-\frac{1}{2} 
\sum_{n=1}^{N} m_{n}\left( \bm{\omega} \cdot
\left( \bm{a}_{n}^{0}+ \bm{d}_{n}\right) \right)^{2}   \nonumber \\
&+&  \bm{\omega} \cdot
\sum_{n=1}^{N} m_{n} \left(\bm{d}_{n}\times \frac{d \bm{d}_{n}}{dt}\right)   \nonumber \\
&+& \frac{1}{2}
\sum_{n=1}^{N} m_{n}\frac{d \bm{d}_{n}}{dt}\cdot 
\frac{d \bm{d}_{n}}{dt}.  \label{classickeckart}
\end{eqnarray}
\end{widetext}
That is the translational motion becomes completely
separated from rotation and vibrations, whereas 
the rotational and vibrational motions are decoupled at the reference 
configuration (at zero displacements). Eqs. (\ref{reckart})  
are called the rotational Eckart conditions. The translational and rotational Eckart conditions together
are referred to as Eckart conditions. 

Displacements obeying the Eckart conditions will be  called Eckart displacements and denoted as
$\bm{d}_{n}^{\textrm{E}}$.  
Atom coordinates $\bm{a}_{n}=\left( a_{xn},a_{yn},a_{zn}\right)^{T}$ obeying equations 
similar to the Eckart conditions,
\begin{gather}
\sum_{n=1}^{N} m_{n} \bm{a}_{n} =0, \\
\sum_{n=1}^{N} m_{n} \bm{a}_{n}^{0} \times \bm{a}_{n} =0,
\end{gather}
will be called Eckart coordinates and denoted as $\bm{a}_{n}^{\textrm{E}}$.

Clearly, 
to derive an Eckart KEO one must calculate Eckart displacements (or Eckart coordinates).

\subsection{Eckart displacements and Eckart coordinates}

It should be clear from Subsection\, \ref{eckartcond} that it is the Eckart conditions and not
some of their derivatives, the Eckart-axis conditions for example, which enter directly 
into the derivation of the nuclear motion energy operator. Therefore, we must look for solutions of the Eckart conditions. 
By definition $\bm{d}_{n}^{\textrm{E}}$ are such solutions. In other words, they solve the
system of equations, a homogeneous system of six linear equations with $3N$ unknowns, provided by the Eckart conditions: 
\begin{gather}
\bm{E}\bm{d}=0,
\end{gather}
where
\begin{gather}
\bm{d}^{T}
=\left( \bm{d}_{1}^{T},\bm{d}_{2}^{T},\ldots, \bm{d}_{N}^{T} \right),
\end{gather}
and 
matrix $\bm{E}$ can be read out from the Eckart conditions.

The general solution of such a system of equations\cite{carl-meyer} is 
\begin{gather}
\bm{d}^{\textrm{E}} = \sum_{j=1}^{K} b_{j}\bm{h}^{\textrm{E},j},
\end{gather}
where $K=3N-6$ (since $\textrm{rank} (\bm{E})=6$), $b_{j}$ are free variables, and $\bm{h}^{\textrm{E},j}$ are particular solutions
(that is $\bm{h}^{\textrm{E},j}$ obey the Eckart conditions).

In Appendix \ref{eckart-sol} analytical expressions of 
$\bm{h}^{\textrm{E},j}$ applicable to any molecule are derived. They depend only on the 
atomic masses and the atom coordinates of the reference configuration. 
Though it is less general than the method described in Appendix \ref{eckart-sol},
it is interesting to note that particular solutions
of the Eckart conditions can be also obtained analytically by choosing $3N-6~ (3N-5)$ internal coordinates and
calculating the corresponding Wilson $\bm{s}_{j}$ vectors\cite{wilson}. Due to the translational and rotational invariance
of the $\bm{s}_{j}$ vectors calculated at the reference configuration the vectors 
$\bm{h}^{\textrm{E},j}= \bm{m}^{-1}\bm{s}_{j}$
are particular solutions, where
$\bm{m}$ denotes
a $3N$ by $3N$ diagonal matrix with
\begin{gather}
\textrm{diag} (\bm{m})
= \left(
\begin{array}{cccccccccc}
m_{1} & m_{1} & m_{1} & m_{2} & m_{2} & m_{2} & \ldots & m_{N} & m_{N} & m_{N}
\end{array}
\right). \nonumber
\end{gather}

The vectors $\bm{h}^{\textrm{E},j}$ span a subspace of the 
configuration space. It is called vibrational space.
The vectors $\bm{h}^{\textrm{E},j}$ are, in general, not orthogonal.
To simplify calculations we shall use in place of $\bm{h}^{\textrm{E},j}$ a set of vectors 
$\bm{d}^{\textrm{E},j}$
obeying the
normalization condition 
\begin{gather}
\left[ {\bm{d}}^{\textrm{E},j} \right]^{T}\bm{m}{\bm{d}}^{\textrm{E},i} = \delta_{ji},
\end{gather}
and write the general solution of the Eckart conditions as
\begin{gather}
{\bm{d}}_{n}^{\textrm{E}} = \sum_{j=1}^{3N-6} c_{j} {\bm{d}}_{n}^{\textrm{E},j}
\end{gather}
with $c_{j}$ denoting free parameters.

Mathematically the $c_{j}$ are just free parameters. Physically, they are vibrational coordinates. The 
relations of the LS coordinates and nuclear motion coordinates
assuming the Eckart conditions read as
\begin{gather} 
u_{\overline{\alpha}n} =
R_{\overline{\alpha}}
+ \left[ \bm{S}^{-1}\left( \theta,\phi,\chi\right)\right]_{\overline{\alpha}\beta}\bm{a}_{\beta n}^{\textrm{E}}, \\
\bm{a}_{\beta n}^{\textrm{E}}=
\bm{a}_{\beta n}^{0}+\bm{d}^{\textrm{E}}_{\beta n},
\\
\bm{d}^{\textrm{E}}_{\beta n} =
\sum_{j=1}^{3N-6} c_{j}d^{\textrm{E},j}_{\beta n},
\end{gather}
and now, one can derive the nuclear motion Hamiltonian
corresponding to the Eckart conditions. The derivation gives  
the gateway Hamilton operator \cite{szalay-gateway}.
However, the question arises if one can use vibrational coordinates other than the $c_{j}$s.

\subsection{Vibrational coordinate dependence of Eckart displacements and coordinates}

The $c_{j}$ coefficients have geometrical significance.
They are the coordinates of mass-weighted Eckart
displacements, $\bm{m}^{1/2}\bm{d}^{\textrm{E}}$, in a coordinate system
of $3N-6$ orthogonal axes in the space
of $3N$-dimensional vectors satisfying the Eckart conditions.
That is
\begin{gather}
c_{j} = \left[ \bm{m}^{1/2}\bm{d}^{\textrm{E},j} \right]^{T} \bm{m}^{1/2}\bm{d}^{\textrm{E}}. \label{defcj}
\end{gather}
Yet, in other words, they are the coordinates of the projection
of a mass-weighted displacement vector onto the vibrational space. The projection matrix is
\begin{gather}
\bm{P} = \sum_{j} \bm{m}^{1/2}\bm{d}^{\textrm{E},j} \left( \bm{m}^{1/2}\bm{d}^{\textrm{E},j} \right)^{T}. \label{projmat1}
\end{gather}

Therefore, the $c_{j}$ can be expressed in terms of geometrically defined coordinates.

In fact, $c_{j}$ can be expressed in terms of any sets of internal coordinates,
$\left\{s_{r}\right\}$, as follows:
\begin{itemize}
	\item Express the atom coordinates, $b_{\tilde{\alpha} n}$, 
where $\tilde{\alpha}=\texttt{X,Y,Z}$ label coordinates in an initial Cartesian system of axes,
	as functions of the chosen set of internal coordinates,
i.e. $b_{\tilde{\alpha} n}=b_{\tilde{\alpha} n}\left( \left\{s_{r}\right\}\right)$.
	How this may be done has been studied in a number of papers \cite{thompson,hilderbrandt,lopata-kiss,trigub,essen}. 
Then translate to the center of mass
to obtain the coordinates $\bm{a}\left( \left\{s_{r}\right\}\right)$. 
	\item Form the displacement vector $\bm{d}\left( \left\{s_{r}\right\}\right)
	= \bm{a}\left( \left\{s_{r}\right\}\right)-\bm{a}^{0}$.
	\item Project it to obtain the Eckart displacement vector
	$\bm{m}^{1/2}\bm{d}^{\textrm{E}}\left( \left\{s_{r}\right\}\right)$:
	\begin{gather}
	\bm{m}^{1/2}\bm{d}^{\textrm{E}}\left( \left\{s_{r}\right\}\right)
	=\bm{P}\bm{m}^{1/2}\bm{d}\left( \left\{s_{r}\right\}\right).
	\end{gather}
	\item Then it follows, that
	\begin{eqnarray}
	c_{j}\left( \left\{s_{r}\right\}\right) & = &\left[ \bm{m}^{1/2}\bm{d}^{\textrm{E},j}\right]^{T}
	\bm{m}^{1/2}\bm{d}^{\textrm{E}}\left( \left\{s_{r}\right\}\right) \nonumber \\
	& = & \left[ \bm{m}^{1/2}\bm{d}^{\textrm{E},j}\right]^{T} \bm{m}^{1/2}\bm{a}\left( \left\{s_{r}\right\}\right)-c^{0}_{j},
	\end{eqnarray}
	where
	\begin{gather}
	c^{0}_{j} = \left[ \bm{m}^{1/2}\bm{d}^{\textrm{E},j}\right]^{T} \bm{m}^{1/2}\bm{a}^{0}.
	\end{gather}
\end{itemize}

\subsection{The general-form Eckart ro-vibrational Hamilton operator} \label{gen-eck-h}

The gateway Hamilton operator is an Eckart ro-vibrational Hamiltonian with exact KEO given in terms 
of the vibrational coordinates $c_{j}$. It is called gateway, since, 
due to the simple relation of the coordinates $c_{j}$ with other sets of vibrational coordinates, 
it can be easily transformed into Eckart Hamiltonians given in terms of other vibrational coordinates.
The various terms in the general-form Eckart Hamilton operator given in terms of a general set of vibrational
coordinates $s_{r}$, as obtained by transforming the gateway Hamiltonian, are summarized
in Tables\, \ref{tab:rotkin}, \ref{tab:vibkin}, \ref{tab:rovibkin}, and \ref{tab:poten}.

\begin{table*}[!htb] 
\caption{\label{tab:rotkin}The rotational KEO }
\begin{ruledtabular}
\begin{tabular}{@{\hspace{3em}} l|c @{\hspace{3em}}}
\scalebox{0.9}{$2\hat{T}_{\textrm{rot}}$}  &   \scalebox{0.9}{$\hat{J}_{\alpha}\mu_{\alpha\beta}\hat{J}_{\beta}$}  \\
\hline
\scalebox{0.9}{$\mu_{\alpha\beta}$} & \scalebox{0.9}{$\left(\bm{I}^{'-1}\right)_{\alpha\beta}$} \\
\scalebox{0.9}{$I^{'}_{\alpha\beta}$} & 
\scalebox{0.9}{$~~I_{\alpha\beta}-\sum\limits_{jkl} \zeta^{\alpha}_{jl}\zeta^{\beta}_{kl}c_{j}c_{k}
=I^{''}_{\alpha\gamma} \left( \left[ \bm{I}^{0}\right]^{-1}\right)_{\gamma\delta} I^{''}_{\delta\beta}$} \\
\scalebox{0.9}{$\zeta^{\alpha}_{jl}$} & 
\scalebox{0.9}{$\epsilon_{\alpha\beta\gamma}\sum\limits_{n} m_{n}d_{\beta n}^{\textrm{E},j} 
d_{\gamma n}^{\textrm{E},l}$} \\
\scalebox{0.9}{$I^{''}_{\alpha\beta}$} & \scalebox{0.9}{$I^{0}_{\alpha\beta} 
+ \frac{1}{2} \sum\limits_{k} a_{k}^{\alpha\beta}c_{k}$} \\
\scalebox{0.9}{$a_{k}^{\alpha\beta}$} & 
\scalebox{0.9}{$2\epsilon_{\alpha\gamma\epsilon}\epsilon_{\beta\delta\epsilon} \sum\limits_{n} 
m_{n} a_{\gamma n}^{0}d_{\delta n}^{\textrm{E},k}$}
\end{tabular}
\end{ruledtabular}
\end{table*}    
      
\begin{table*}[!htb] 
\caption{The vibrational KEO}\label{tab:vibkin}
\begin{ruledtabular}
\begin{tabular}{@{\hspace{3em}}l|c@{\hspace{3em}}}
\scalebox{0.9}{$2\hat{T}_{\textrm{vib}}$} & 
\scalebox{0.9}{$\varrho^{2}\sum\limits_{r} \hat{p}_{r} 
\mathcal{G}_{rs}\hat{p}_{s}$}  \\
\hline
\scalebox{0.9}{$\mathcal{G}_{rs}$} &  \scalebox{0.9}{$G_{rs}+\varrho^{2} C_{\alpha r}\mu_{\alpha\beta}C_{\beta s}$} \\
\scalebox{0.9}{$G_{rs}$}   &  
\scalebox{0.9}{$\sum\limits_{i} \frac{\partial s_{r}}{\partial c_{i}}\frac{\partial s_{s}}{\partial c_{i}}$} \\
\scalebox{0.9}{$C_{\alpha r}$} & 
\scalebox{0.9}{$\sum\limits_{ij}  \zeta^{\alpha}_{ij} c_{i}\frac{ \partial s_{r}}{\partial c_{j}}$} 
\end{tabular}
\end{ruledtabular}
\end{table*}
        
\begin{table*}[!htb] 
\caption{The rotational-vibrational coupling KEO}\label{tab:rovibkin}
\begin{ruledtabular}
\begin{tabular}{@{\hspace{3em}}l|c@{\hspace{3em}}}
 $2\hat{T}_{\textrm{rot-vib}}$    & 
$-\varrho^{2}\sum\limits_{r} \left( \hat{J}_{\alpha}\mathcal{C}_{\alpha r} \hat{p}_{r}
+ \hat{p}_{r} \mathcal{C}^{T}_{r\alpha}\hat{J}_{\alpha} \right)$ \\
\hline
$\mathcal{C}_{\alpha r}$ & $\mu_{\alpha\beta}C_{\beta r}$
\end{tabular}
\end{ruledtabular}
\end{table*}
 	
\begin{table*}[!htb] 
\caption{The potential and the pseudo-potential energy}\label{tab:poten}
\def\arraystretch{1} 
\begin{ruledtabular}
\begin{tabular}{@{\hspace{3em}}l|c@{\hspace{3em}}}
 $V$    & 
\scalebox{0.9}{$V\left( \left\{ a_{\beta n}^{0}+\sum\limits_{j}c_{j}d_{\beta n}^{\textrm{E},j} \right\} \right)$} \\
$V_{\textrm{ps}}$ & 
$\tilde{V}-\frac{\hbar^{2}}{8} \mu_{\alpha\alpha} $ \\
 $\frac{8}{\hbar^{2}}\tilde{V}$ & 	\scalebox{0.9}{$
 -\sum_{ij}\left[ \frac{\partial  {\tilde{\mathcal{G}} }_{ij}}{\partial c_{i}} \right]
  \frac{ \partial \ln\det \bm{G} }{\partial c_{j} } $} 
\scalebox{0.9}{$ +\sum_{ij} {\tilde{\mathcal{G}} }_{ij}
\left\{ 
-\frac{\partial^{2}  \ln\det \bm{G} }{\partial c_{i} \partial c_{j}} + \frac{1}{4} 
\frac{ \partial 
\ln\det \bm{G} }{\partial c_{i} }
\frac{ \partial 
\ln\det \bm{G} }{\partial c_{j} }
\right\} $} \\
& 
\scalebox{0.9}{$
-\sum_{ij}\sum_{r} \left[ \frac{ \partial}{\partial s_{r}} \frac{\partial s_{r}}{\partial c_{i}}\right] 
{\tilde{\mathcal{G}} }_{ij} \frac{ \partial \ln \left( \det\bm{\mu} \det\bm{G}\right)}{\partial c_{j}}
$} 
\scalebox{0.9}{$
-\sum_{ijk}\sum_{s}  
\left[ \frac{ \partial}{\partial c_{i}} \frac{\partial s_{s}}{\partial c_{j}}\right]
\frac{\partial c_{k}}{\partial s_{s}} \frac{\partial \ln \left( \det\bm{\mu} \det\bm{G}\right)}{\partial c_{k}}
$} \\
&  
\scalebox{0.9}{$
+\frac{1}{4}\sum_{ij} {\tilde{\mathcal{G}} }_{ij}
\left\{
\frac{ \partial \ln\det\bm{\mu}}{\partial c_{i}} 
\frac{ \partial \ln\det\bm{G}}{\partial c_{j}} 
+\frac{ \partial \ln\det\bm{G}}{\partial c_{i}}\frac{ \partial \ln\det\bm{\mu}}{\partial c_{j}}
\right\}
$} \\
${\tilde{\mathcal{G}} }_{ij}$ & \scalebox{0.9}{$ \delta_{ij} +\sum_{kl} c_{k} \zeta^{\alpha}_{ki} \mu_{\alpha\beta}
\zeta^{\beta}_{lj}c_{l}
$}  
\end{tabular}
\end{ruledtabular}
\end{table*}

By examining the expressions one can see that no derivatives have to be calculated to obtain the rotational matrix 
$\bm{\mu}$, but one has to calculate the derivatives of the coordinates $s$ with respect to the coordinates $c$
to evaluate the vibrational matrix $\mathcal{G}$.
One must also note that the vibrational matrix has contribution from the rotational matrix $\bm{\mu}$ and the matrix
$\bm{C}$ appearing in the terms coupling rotational and vibrational momenta. Therefore, to check the validity and demonstrate
the usefulness of the gateway method it suffices to consider the vibrational part of the general-form Eckart Hamiltonian.

In passing, it should be noted that the term $\tilde{V}$ was left out from the expression
of the pseudo-potential in Ref.\, \onlinecite{szalay-gateway}. It was assumed that the
pseudo-potential transformed as an ordinary scalar function under change of coordinates. 
But this is not a valid assumption \cite{littlejohn-reinsch}. $\tilde{V}$ was also omitted in 
Ref.\, \onlinecite{szalay-aspects}, since
it was assumed that Eq. (71) of Ref.\, \onlinecite{szalay-aspects} held. This equation does not hold in general, however. 

\section{Solution of the vibrational Schr\"odinger equation}

Numerical implementation of constructing the gateway vibrational Hamilton operator and a method of solving the
corresponding Schr\"odinger equation is described below.

The vibrational Schr\"odinger equation reads as
\begin{gather}
\left[ \frac{1}{2}\sum_{rs} \hat{p}_{r} \mathcal{G}_{rs}\hat{p}_{s} 
+V +V_{\textrm{ps}}\right] \vert \Psi \rangle = \lambda \vert \Psi \rangle ,
\nonumber 
\end{gather}  
where $\hat{p}_{r}=-i\hbar\partial/\partial s_{r}$.
When solving this equation we do not aim at extreme accuracy nor we want to obtain all vibrational energy levels. Our goal
is just to show that the gateway method works fine. Then, if required, one can improve the results  by employing 
more sophisticated methods to solve the eigenvalue equation. Therefore, we choose as simple a method of solution as possible:
\begin{itemize}
\item Direct product {\it sinc} discrete variable representation (DVR)\cite{colbert-miller,szalay-smith,littlejohn-cargo,jerke}, 
a special case of discretized continuous contracted Hermite distributed approximating functions \cite{szalay-smith}, is employed.  
\item Approximate eigenvalues are obtained by Lanczos iteration\cite{lanczos,cullum}. 
\item The calculation of a Hamiltonian matrix vector product is carried out by partial summation
\cite{bramley-carrington}. It
 scales as $n^{3N-5}$ where $n$ is the number
of basis functions applied to a single vibrational mode. (The Hamiltonian matrix is never calculated.)
\end{itemize}

The calculations consist of two main steps:
\begin{itemize}
\item The elements of the kinetic energy matrix and the potential energy are calculated on a grid. 
\item The Lanczos algorithm is employed to obtain approximate energy levels.  
\end{itemize}

The first step is split. At first quantities independent of the grid points shown in Figure 1 are calculated.
\begin{figure*}[!htb]
\begin{center}
\includegraphics[scale=0.8,keepaspectratio]{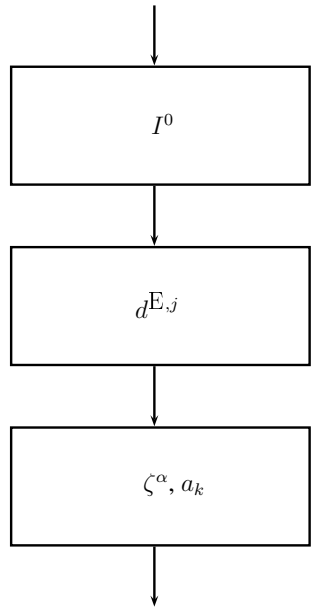}
\caption{Evaluation of the vibrational Hamilton operator: Quantities independent of the grid points. \\}
\end{center}
\end{figure*}
Then the calculation continues by calculating the vibrational kinetic matrix and the potential and
pseudo-potential values at the grid points as shown in Figure 2.
\begin{figure*}[!htb]
\begin{center}
\includegraphics[scale=0.8,keepaspectratio]{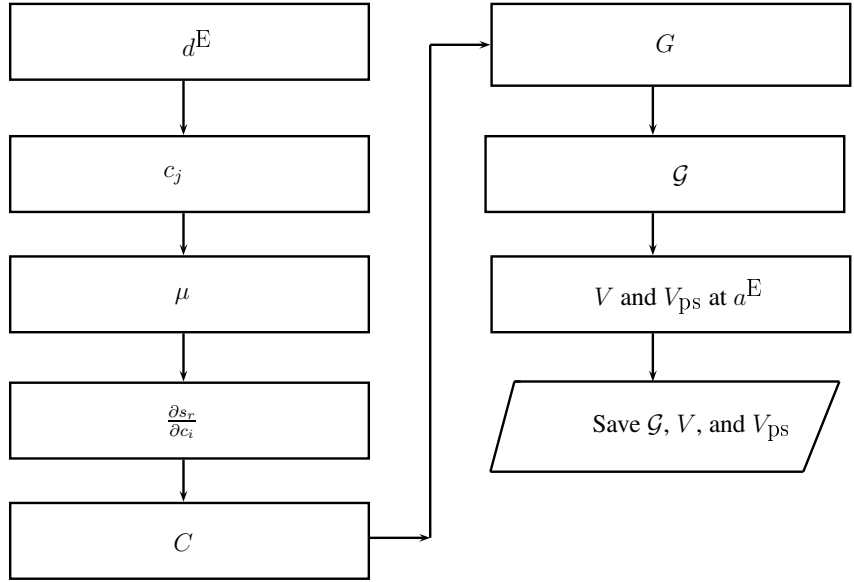} \\
\caption{Evaluation of the vibrational Hamilton operator: Grid dependent quantities.\\}
\end{center}
\end{figure*}
The difficult looking steps are the calculations of the derivatives $\frac{\partial s_{r}}{\partial c{j}}$ and the pseudo-potential.
Even these can be done quite simply. Note that:
\begin{itemize}
\item When Eckart displacements are obtained by rotation,
\begin{gather}
 \frac{\partial s_{r}}{\partial c_{j}}  =  \frac{\partial s_{r}\left( \bm{a}\right)}{\partial c_{j}}=
\frac{\partial s_{r}\left( \bm{a}^{\textrm{E}}\right)}{\partial c_{j}} =
 \frac{\partial s_{r}}{\partial \bm{a}^{\textrm{E}} }
 \frac{\partial \bm{a}^{\textrm{E}}}{\partial c_{j} } \nonumber \\
 = \left. \frac{\partial s_{r}}{\partial \bm{a}}\right|_{\bm{a}=\bm{a}^{\textrm{E}}} \bm{d}^{\textrm{E},j}
= \bm{s}_{r}(\bm{a}^{\textrm{E}}) \cdot  \bm{d}^{\textrm{E},j},
\end{gather}
where $\bm{s}_{r}(\bm{a}^{\textrm{E}})$ is the Wilson $\bm{s}$-vector corresponding 
to the $r$th vibrational coordinate calculated at
the $\bm{a}^{\textrm{E}}$ configuration. Note that the 
calculation of $\frac{\partial s_{r}}{\partial \bm{a}}$ is analytical, and
clearly, there is no need for differentiating the matrix rotating into the Eckart frame.
\item  When Eckart displacements are obtained by projection the derivatives 
$\frac{\partial c_{j}}{\partial s_{r}}$ can be obtained analytically 
\begin{gather}
\frac{ \partial c_{j}}{\partial s_{r}} = 
 \left[ \bm{d}^{\textrm{E},j}\right]^{T} \bm{m}
 \frac{\partial \bm{a}\left( \left\{s_{r}\right\}\right)}{\partial s_{r}}.
\end{gather}
Then, the desired derivatives are obtained by inverting a matrix $\bm{D}$ whose elements
are $D_{jr}=\frac{\partial c_{j}}{\partial s_{r}}$.
\end{itemize}

One might be interested in calculating the derivatives of Eckart coordinates. Numerical 
Eckart codes\cite{wang-carrington,szidarovszky-fabri-csaszar,sadri,yachmenev-yurchenko} 
do, in fact, calculate such derivatives.
Since 
\begin{gather}
\bm{a}^{\textrm{E}}= \bm{a}^{0}+\sum_{j=1}^{3N-6}c_{j}\bm{d}^{\textrm{E},j},
\end{gather}
it follows that
\begin{gather}
\frac{\partial \bm{a}^{\textrm{E}}}{\partial s_{r}} = 
\sum_{j=1}^{3N-6}\frac{ \partial c_{j}}{\partial s_{r}}\bm{d}^{\textrm{E},j}. 
\end{gather}
Therefore, when projection is employed to calculate Eckart displacements the calculation of the
derivatives of Eckart coordinates with respect to the vibrational coordinates is trivial.
When one employs rotation to generate Eckart displacements one can calculate the
derivatives $\frac{\partial s_{r}}{\partial c_{j}}$ simply by the method described above. Then, the
derivatives $\frac{\partial c_{j}}{\partial s_{r}}$ are obtained by inverting a matrix $\bm{A}$ whose
elements are defined as $A_{rj}=\frac{\partial s_{r}}{\partial c_{j}}$. 
As the first of the numerical examples
demonstrating the working of the gateway method we shall compare this method of differentiating Eckart
coordinates with the simplest numerical differentiation scheme,
\begin{widetext}
\begin{eqnarray}
\text{\scalebox{0.7}{$ 
\left( \frac{\partial \bm{a}_{n}^{\textrm{E}}}{\partial s_{r}}\right)_{\bm{s}=\bm{s}^{0}} $ }} & \approx &
\text{\scalebox{0.7}{$  
\frac{ \bm{U}\left( s_{1}^{0},s_{2}^{0},\ldots,s_{r}^{0}+h,s_{r+1}^{0},\ldots,s_{3N-6}^{0}\right)  
\bm{a}_{n} \left( s_{1}^{0},s_{2}^{0},\ldots,s_{r}^{0}+h,s_{r+1}^{0},\ldots,s_{3N-6}^{0}\right) 
}{2h} \nonumber $ }} \\
&-&\text{\scalebox{0.7}{$\frac{
\bm{U}\left( s_{1}^{0},s_{2}^{0},\ldots,s_{r}^{0}-h,s_{r+1}^{0},\ldots,s_{3N-6}^{0}\right)
\bm{a}_{n} \left( s_{1}^{0},s_{2}^{0},\ldots,s_{r}^{0}-h,s_{r+1}^{0},\ldots,s_{3N-6}^{0}\right) }{2h},
$ }} 
\end{eqnarray}
\end{widetext}
where $h$ is small distortion along a vibrational coordinate, and 
$\bm{U}$ rotates the coordinates of a given distorted configuration into Eckart coordinates.   

The higher order derivatives required to evaluating the pseudo-potential can be obtained similarly, without resorting to numerical differentiation.   

\section{Numerical examples}

All results of numerical calculations to be presented refer to the H$_{2}$O molecule.
The equilibrium geometry and the potential energy surface 
obtained by Jensen by fitting to experimental data \cite{jensen} are used in the calculations.
Valence internal coordinates, the two bond lengths, $r_{1}$ and $r_{2}$, and the bond angle
$\phi$ are used as vibrational coordinates.

\begin{figure*}[!htb]
\begin{center}
\includegraphics[scale=0.8,keepaspectratio]{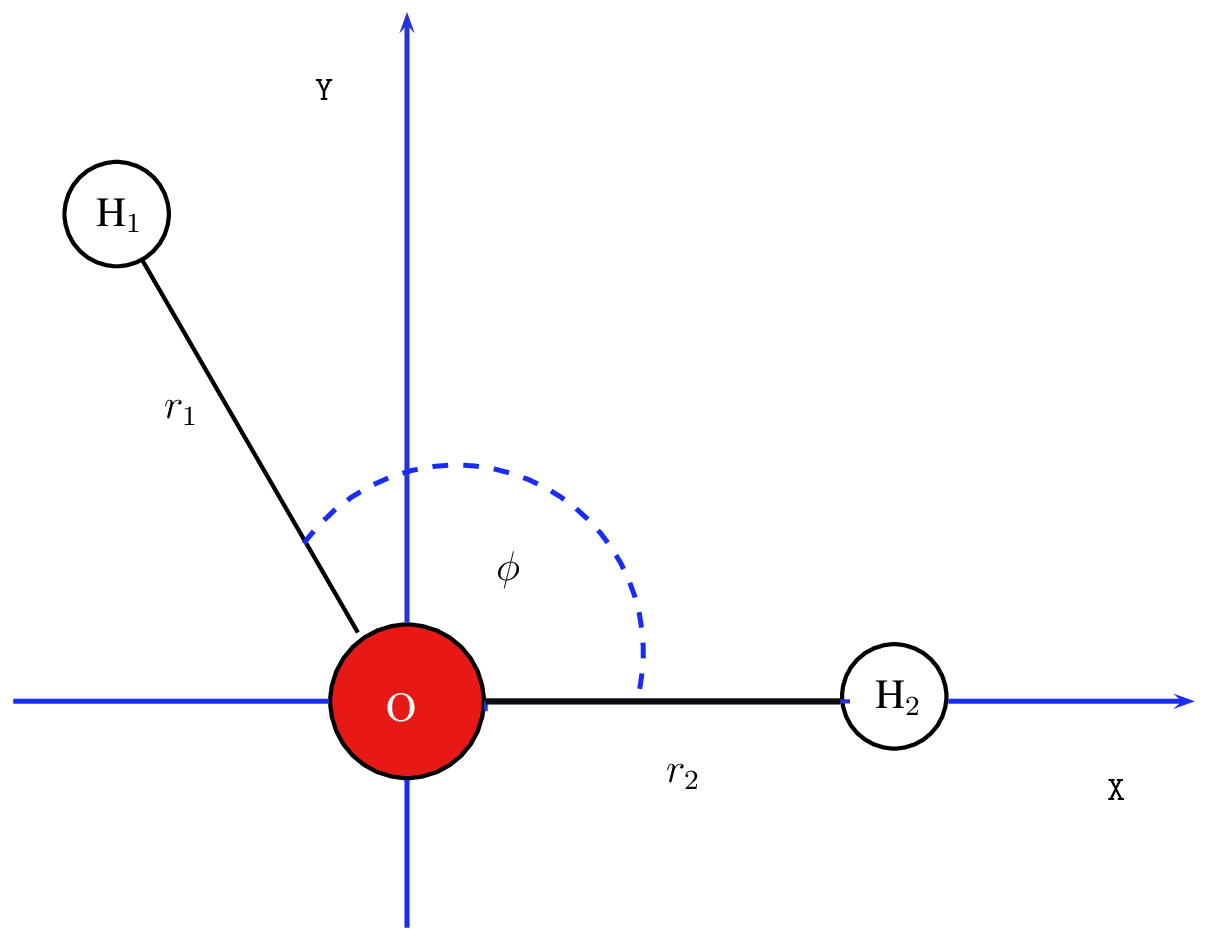}
\caption{Location of the molecule in the initial system of axes ({\tt X,Y,Z)}).} \label{fig:initialaxes}
\end{center}
\end{figure*}

The origin of the MS (Eckart-frame) is fixed to center of mass of the reference 
configuration and its axes are parallel to the axes of the coordinate system, the initial
axis-system, shown in Figure \ref{fig:initialaxes}.

$\bm{a}\left( r_{1},r_{2},\phi \right)$ 
are the coordinates 
of the atoms of 
distorted configurations whose center of mass coincides with the origin of the MS.
They can be 
obtained from the atom coordinates summarized in Table\, \ref{tab:atomcoords}
by translation.
They are used along with the coordinates 
$\bm{a}^{0}=\bm{a}\left( r_{1}^{0},r_{2}^{0},\phi^{0} \right)$
of the reference configuration, where the equilibrium bond lengths, $r_{1}^{(e)}=r_{2}^{(e)}=0.95843~\text{\AA}$, and the 
equilibrium bond angle,  $\phi^{(e)}=104.43976^{\circ}$, 
are taken as $r_{1}^{0},r_{2}^{0}$ and $\phi^{0}$,
respectively,
to calculate Eckart displacements and Eckart coordinates. To obtain 
 Eckart displacements (and Eckart coordinates) 
 we use the rotation and projection methods.
 The matrix $\bm{U}$ rotating to Eckart coordinates is obtained by using 
the method of Ref.\, \onlinecite{krasnos}. The basis vectors of the vibrational space given in Table \ref{tab:vibspacebasis}
are obtained by employing the analytical formulas derived in Appendix \ref{eckart-sol}. 

\begin{table*}[!htb]
\begin{threeparttable}
\caption{Numbering of the atoms, the atomic masses, and the coordinates $b_{\tilde{\alpha}n}$ of 
the atoms in the initial system of axes.} \label{tab:atomcoords}
\begin{ruledtabular}
\begin{tabular}{lccccc}
Atom &  Numbering of the atoms; $n$ & Masses &{\tt X}   & {\tt Y}  & {\tt Z} \\
\hline
H$_{1}$  &  $1$ & $1.00782522$ & $r_{1}\cos\phi$ &  $r_{1}\sin\phi$ & $0$ \\
O        &  $2$ & $15.99491502$ & $0$             & $0$              & $0$\\
H$_{2}$   & $3$ & $1.00782522$ & $r_{2}$         & $0$              &  $0$  
\end{tabular}
\end{ruledtabular}
\begin{tablenotes}
      \small
      \item a) Tha atomic masses are in atomic mass units.
    \end{tablenotes}
\end{threeparttable}
\end{table*}

\begin{table*}[!htb]
\begin{threeparttable}
\caption{Basis vectors for the vibrational space} \label{tab:vibspacebasis}
\begin{ruledtabular}
\begin{tabular}{lccc}
 $\alpha n$&  $\bm{d}^{\textrm{E},1}$   & $\bm{d}^{\textrm{E},2}$  & $\bm{d}^{\textrm{E},3}$ \\
\hline
   $x1$  & -0.2409179258 &   -0.6669945045   &  0.0308844916 \\
   $y1$  &  0.9356174659 &   -0.1315179013   & -0.0084552498 \\
   $z1$ &  0.0000000000  &  -0.0000000000   & -0.0000000000 \\
   $x2$ &  0.0151800220  &   0.0420267243   &  0.0589871470 \\
   $y2$  & -0.0589524156  &   0.0510523585   & -0.0012189233 \\
   $z2$  &  0.0000000000  &  -0.0000000000   &  0.0000000000 \\
   $x3$  &  0.0000000000 &    0.0000000000   & -0.9670531688 \\
   $y3$  &  0.0000000000  &  -0.6787199452   &  0.0278004437 \\
   $z3$  & 0.0000000000  &   0.0000000000    & -0.0000000000
\end{tabular}
\end{ruledtabular}
\begin{tablenotes}
      \small
      \item a) It turns out that with the chosen MS and reference configuration $v^{x}_{x2}=0$. 
Therefore some of the terms appearing
in the expressions of $\bm{h}^{\textrm{E},j}$ are singular. Rotation of the MS around the $y$ axis
by $\pi/2$ removes the singularities and $\bm{h}^{\textrm{E},j}$ as well as 
$\bm{d}^{\textrm{E},j}$ can be calculated in the rotated MS. Then returning to the MS gives 
gives the desired basis vectors of vibrational space.
    \end{tablenotes}
\end{threeparttable}
\end{table*}
   
\subsection{Differentiation of Eckart coordinates}

Table VII compares the derivatives of Eckart coordinates with respect to the bond angle at a distorted
configuration. 
As expected in comparison of approximate numerical differentiation results with analytical ones, the deviations of the gateway and  
the numerical differentiation results are the smallest 
at an intermediate $h$. Thus, the results by the gateway method do qualify as "analytical (exact)" results.

\begin{table*}[!htb] 
\caption{Derivatives of Eckart coordinates with respect to $\phi$
at $\left( r_{1}=1.358430 \text{\AA},~r_{2}=0.658430 \text{\AA},~\phi=1.322818\textrm{rad}  \right)$.}
\label{derivatives} 
\def\arraystretch{0.8}
\begin{ruledtabular}
\begin{tabular}{cccc}
$h$ & Approximate $\partial \bm{a}^{\textrm{E}}/\partial \phi$ & Exact
 $\partial \bm{a}^{\textrm{E}}/\partial \phi$
& Approx$-$Exact\\
\hline \\
        0.1000000000 &   -0.4054459586 &   -0.4057377040 &    0.0002917454 \\
        0.1000000000 &   -0.0207341020 &   -0.0206473673 &   -0.0000867347 \\
        0.1000000000 &    0.0000000000 &    0.0000000000 &    0.0000000000 \\
        0.1000000000 &    0.0145783041 &    0.0145969975 &   -0.0000186934 \\
        0.1000000000 &    0.0263719843 &    0.0263829584 &   -0.0000109741 \\
        0.1000000000 &    0.0000000000 &    0.0000000000 &    0.0000000000 \\
        0.1000000000 &    0.1740777305 &    0.1740727981 &    0.0000049323 \\
        0.1000000000 &   -0.3978083597 &   -0.3980692607 &    0.0002609009 \\
        0.1000000000 &    0.0000000000 &    0.0000000000 &    0.0000000000 \\
  & & & \\
      0.0000100000 &   -0.4057377040 &   -0.4057377040 &    0.0000000000 \\
        0.0000100000 &   -0.0206473675 &   -0.0206473673 &   -0.0000000002 \\
        0.0000100000 &    0.0000000000 &    0.0000000000 &    0.0000000000 \\
        0.0000100000 &    0.0145969975 &    0.0145969975 &    0.0000000000 \\
        0.0000100000 &    0.0263829584 &    0.0263829584 &    0.0000000000 \\
        0.0000100000 &    0.0000000000 &    0.0000000000 &    0.0000000000 \\
        0.0000100000 &    0.1740727982 &    0.1740727981 &    0.0000000000 \\
        0.0000100000 &   -0.3980692606 &   -0.3980692607 &    0.0000000001 \\
        0.0000100000 &    0.0000000000 &    0.0000000000 &    0.0000000000 \\
  & & & \\
       0.0000000001 &   -0.4057600000 &   -0.4057377040 &   -0.0000222960 \\
        0.0000000001 &   -0.0206500000 &   -0.0206473673 &   -0.0000026327 \\
        0.0000000001 &    0.0000000000 &    0.0000000000 &    0.0000000000 \\
        0.0000000001 &    0.0145995000 &    0.0145969975 &    0.0000025025 \\
        0.0000000001 &    0.0263810000 &    0.0263829584 &   -0.0000019584 \\
        0.0000000001 &    0.0000000000 &    0.0000000000 &    0.0000000000 \\
        0.0000000001 &    0.1740700000 &    0.1740727981 &   -0.0000027981 \\
        0.0000000001 &   -0.3980600000 &   -0.3980692607 &    0.0000092607 \\
        0.0000000001 &    0.0000000000 &    0.0000000000 &    0.0000000000 
\end{tabular}
\end{ruledtabular}
\end{table*}

\subsection{The vibrational Hamiltonians} \label{vibrational-hs}

\begin{figure*}[!htb]
\begin{center}
\includegraphics[scale=0.8,keepaspectratio]{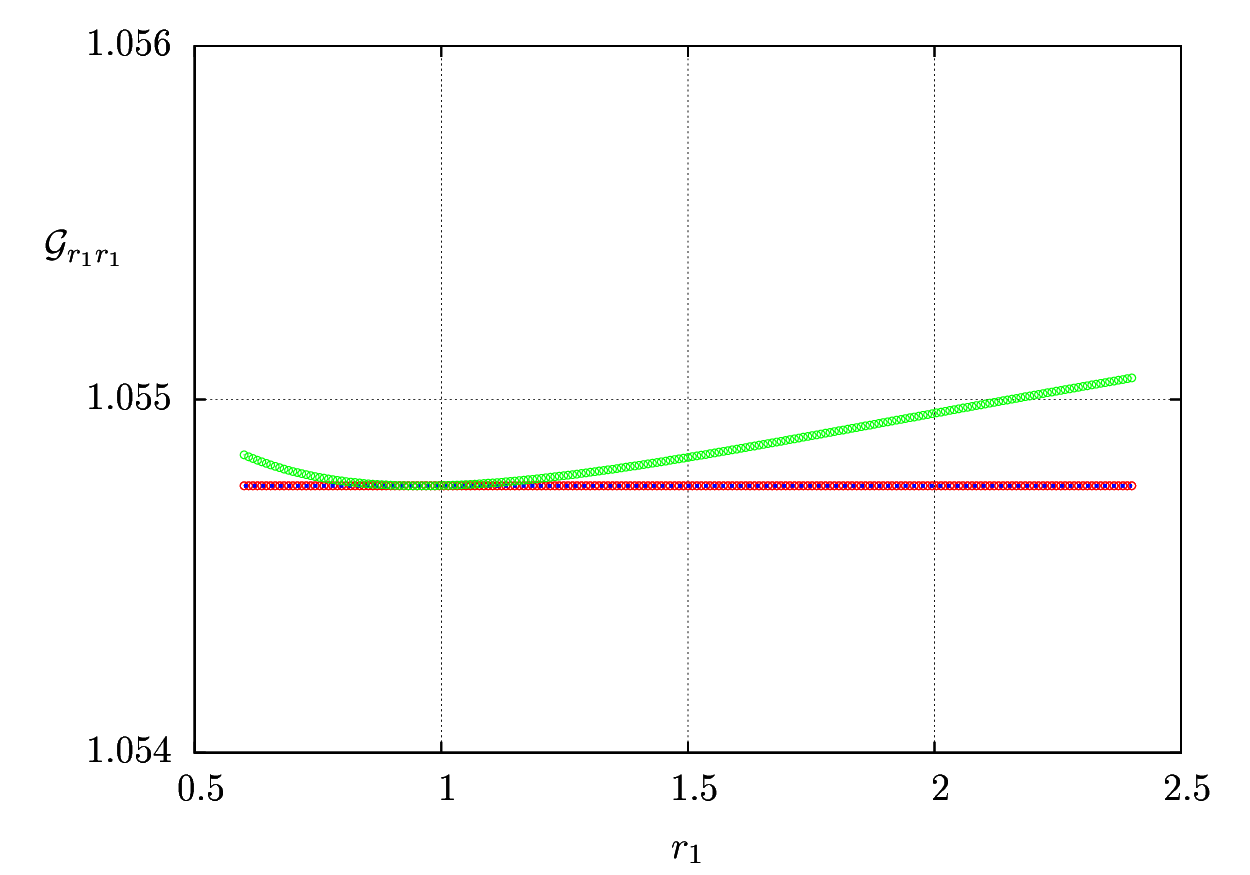}
\caption{$\mathcal{G}_{r_{1}r_{1}}$ as function of  $r_{1}$ when $r_{2}$ and $\phi$
are fixed at their equilibrium values. $\mathcal{G}_{r_{1}r_{1}}$ is in units of $amu^{-1}{\textrm{\AA}}^{-2}$.
Curves in red, green, and blue correspond to the rotation method, the projection method, and 
the Wilson G-matrix element, $G_{r_{1}r_{1}}\left( r_{1} \right)=\frac{1}{m_{\textrm{H}}}+\frac{1}{m_{\textrm{O}}}$, respectively.
} \label{fig:Gr1r1-r1}
\end{center}
\end{figure*}

\begin{figure*}[!htb]
\begin{center}
\includegraphics[scale=0.8,keepaspectratio]{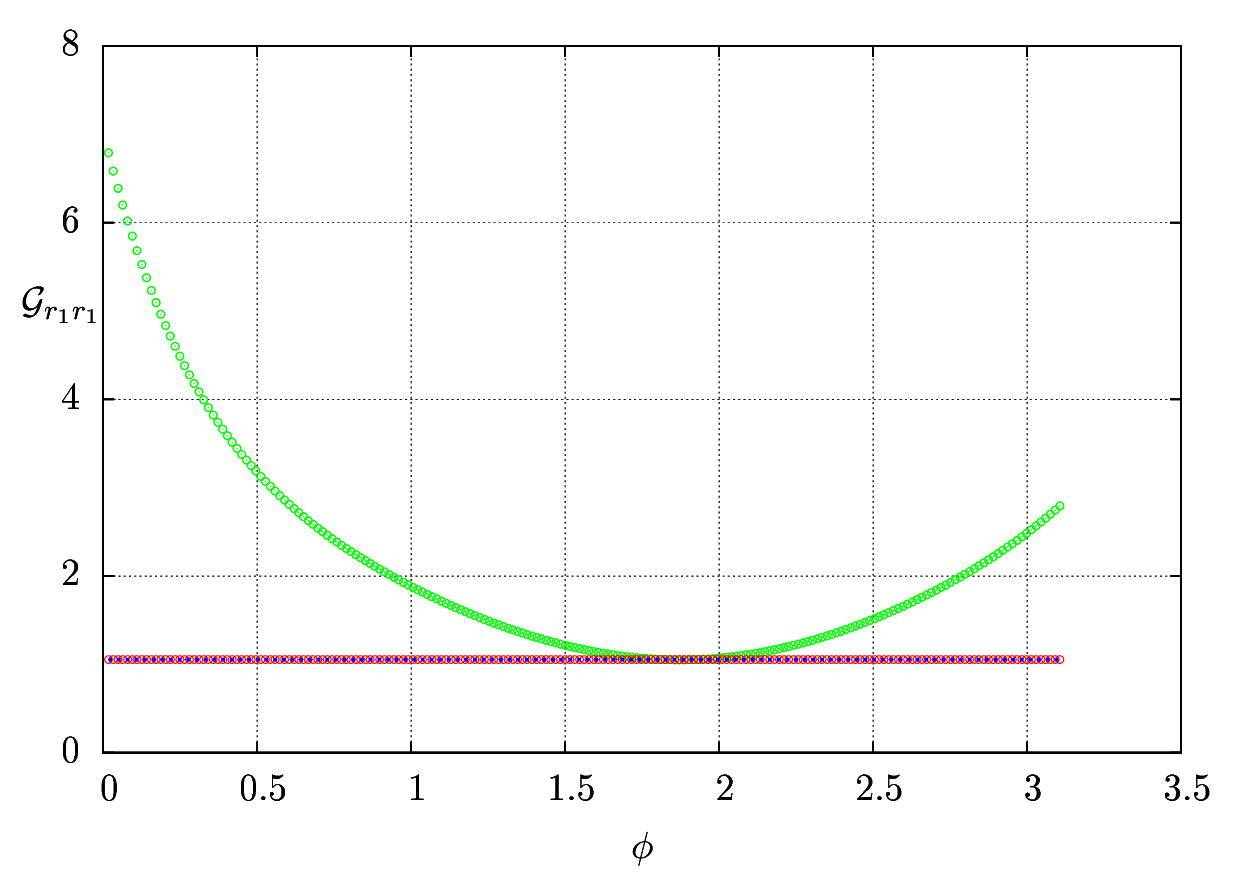}
\caption{$\mathcal{G}_{r_{1}r_{1}}$ as function of the bending angle when $r_{1}$ and $r_{2}$
are fixed at their equilibrium values. $\mathcal{G}_{r_{1}r_{1}}$ is in units of $amu^{-1}{\textrm{\AA}}^{-2}$.
Curves in red, green, and blue correspond to the rotation method, the projection method, and 
the Wilson G matrix element, $G_{r_{1}r_{1}}\left( \phi \right)=\frac{1}{m_{\textrm{H}}}+\frac{1}{m_{\textrm{O}}}$, 
respectively. } \label{fig:Gr1r1-phi}
 \end{center}
\end{figure*}

\begin{figure*}[!htb]
\begin{center}
\includegraphics[scale=0.8,keepaspectratio]{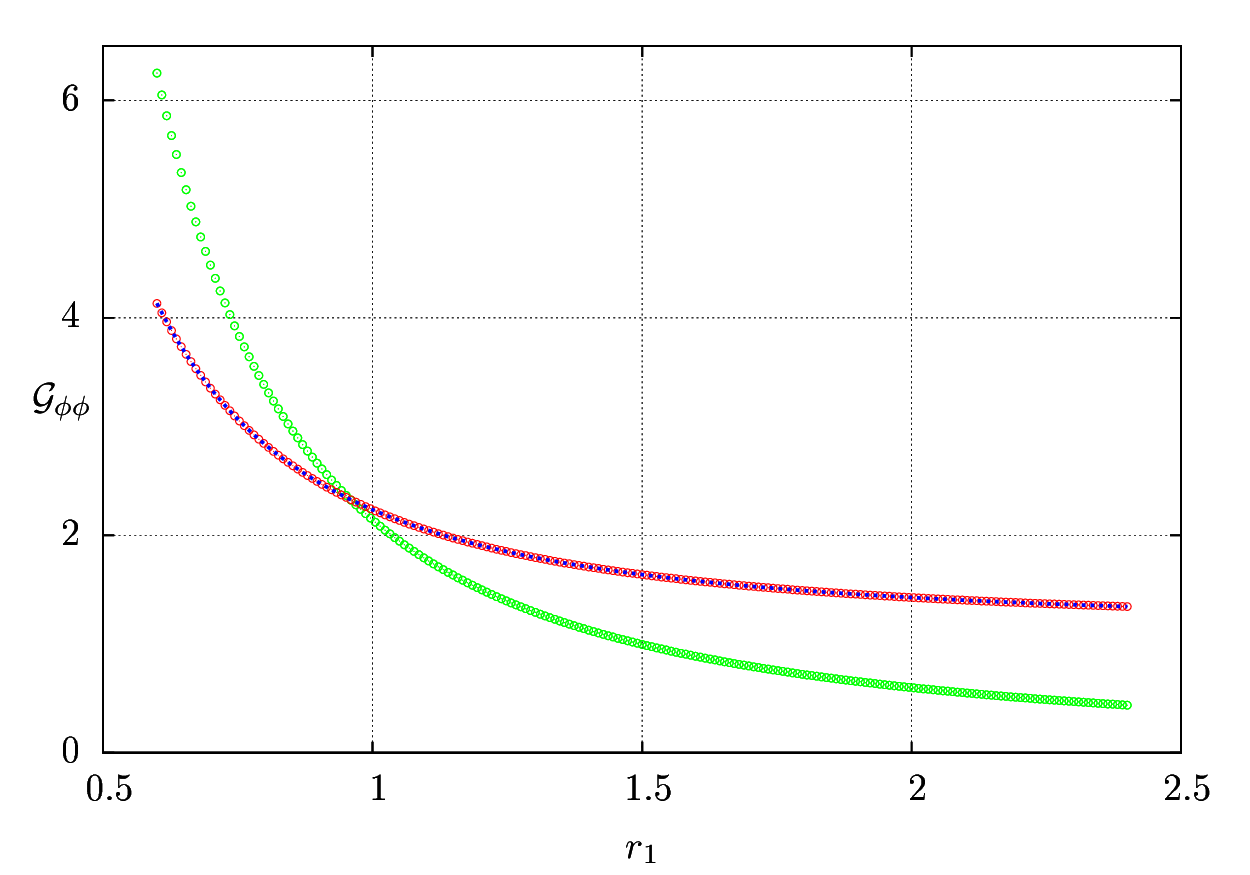}
\caption{$\mathcal{G}_{\phi\phi}$ as function of $r_{1}$  when $r_{2}$ and $\phi$
are fixed at their equilibrium values. $\mathcal{G}_{\phi\phi}$ is in units of $amu^{-1}{\textrm{\AA}}^{-2}$.
Curves in red, green, and blue correspond to the rotation method, the projection method, and 
the appropriate Wilson G matrix element, 
$G_{\phi\phi}\left( r_{1} \right)=\frac{1}{\mu_{\textrm{H}}r_{1}^{2}}
+\frac{1}{\mu_{\textrm{H}}\left(r_{2}^{(e)}\right)^{2}}
-\frac{2\cos\phi^{(e)}}{m_{\textrm{O}}r_{2}^{(e)}r_{1}}
$,  
respectively. ($\mu_{\textrm{H}}=\frac{1}{m_{\textrm{H}}}+\frac{1}{m_{\textrm{O}}}$.) 
} \label{fig:Gphiphi-r1}
\end{center}
\end{figure*}

\begin{figure*}[!htb]
\begin{center}
\includegraphics[scale=0.8,keepaspectratio]{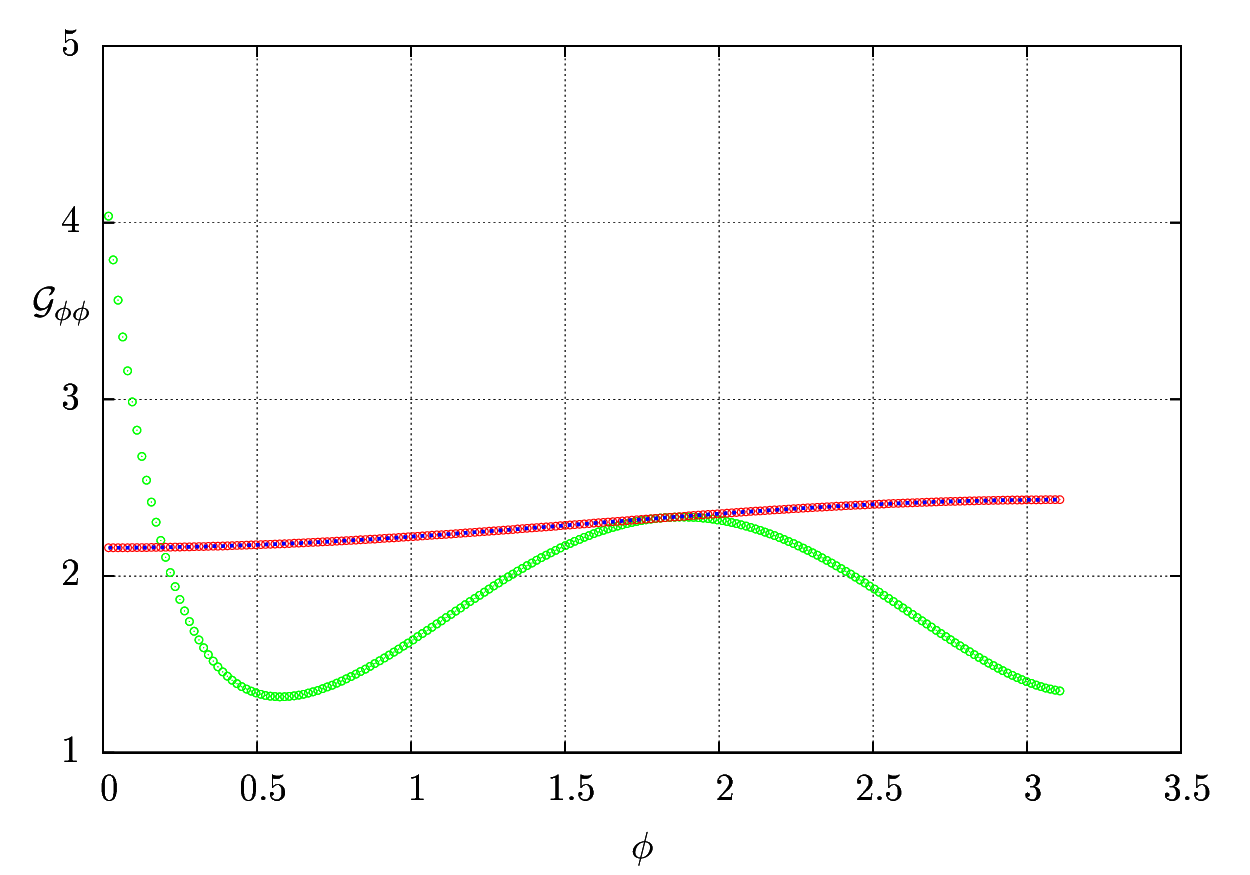}
\caption{$\mathcal{G}_{\phi\phi}$ as function of $\phi$  when $r_{1}$ and $r_{2}$
are fixed at their equilibrium values. $\mathcal{G}_{\phi\phi}$ is in units of $amu^{-1}{\textrm{\AA}}^{-2}$.
Curves in red, green, and blue correspond to the rotation method, the projection method, and 
the appropriate Wilson G matrix element, 
$G_{\phi\phi}\left( \phi \right)=
\frac{1}{\mu_{\textrm{H}}\left( r_{1}^{(e)}\right)^{2}}
+\frac{1}{\mu_{\textrm{H}}\left(r_{2}^{(e)}\right)^{2}}
-\frac{2\cos\phi}{m_{\textrm{O}}r_{2}^{(e)}r_{1}^{(e)}}
$, respectively. ($\mu_{\textrm{H}}=\frac{1}{m_{\textrm{H}}}+\frac{1}{m_{\textrm{O}}}$.)
} \label{fig:Gphiphi-phi}
\end{center}
\end{figure*}

\begin{figure*}[!htb]
\begin{center}
\includegraphics[angle=270,scale=0.6,keepaspectratio]{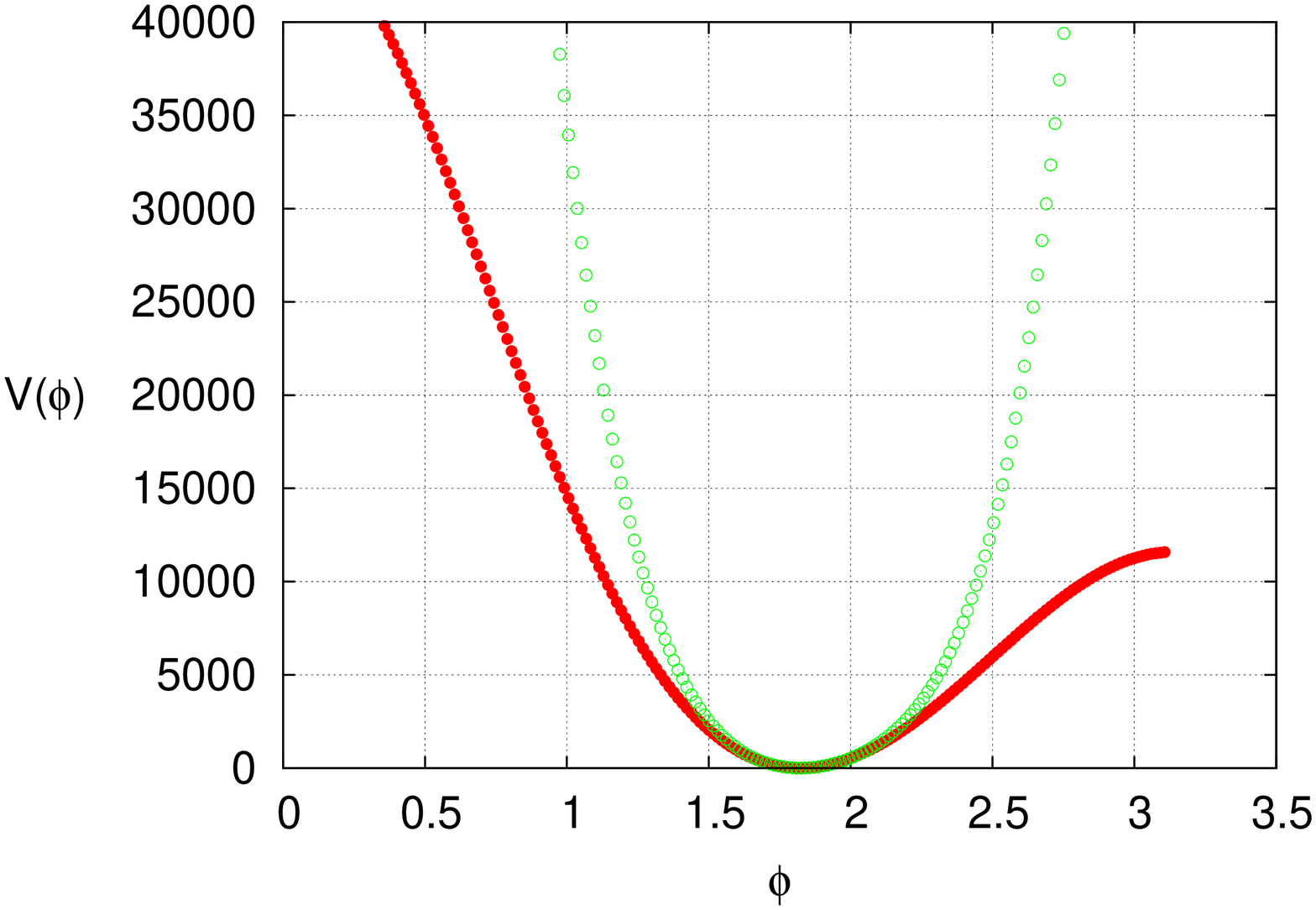}
\caption{The bending potential when the bond lengths are fixed at their
equilibrium values. The red curve is obtained by the rotation method. The green curve is obtained by the projection method.
The potential energy is given in units of cm$^{-1}$.}
\label{fig:bendpes}
\end{center}
\end{figure*}

The potential energy surface and the elements
of the vibrational matrix $\mathcal{G}$ are functions
of the vibrational coordinates. Various 1D cuts
of these surfaces, Figures\, \ref{fig:Gr1r1-r1}, \ref{fig:Gr1r1-phi}, \ref{fig:Gphiphi-r1}, \ref{fig:Gphiphi-phi},
and \ref{fig:bendpes},
 show that the vibrational operators
obtained by employing the rotation and projection methods, respectively, to construct Eckart displacements are different.

The vibrational $\mathcal{G}$ matrix should not depend on rotation to Eckart coordinates. Thus,  
it should be identical with Wilson's G-matrix 
considered as function of the internal coordinates. 
Figures\, \ref{fig:Gr1r1-r1}, \ref{fig:Gr1r1-phi}, \ref{fig:Gphiphi-r1}, and \ref{fig:Gphiphi-phi}
show that this is indeed the case, thus providing another numerical evidence of the correctness of the
gateway method. 

To calculate the pseudo-potential we use the expression
\begin{gather}
V_{\textrm{ps}}=-\frac{\hbar^{2}}{2}\left( \frac{\cos\phi}{r_{1}r_{2}}+\frac{1}{4}G_{\phi\phi}\left( 1+\csc^{2}\phi\right) \right)
\end{gather}
given in Ref.\, \onlinecite{johnson-reinhardt} when Eckart displacements are obtained by the rotation method.
When using projection to obtain Eckart displacements, the pseudo-potential is calculated by employing the formulas
given in Table\, \ref{tab:poten}. Even this complicated expression of the pseudo-potential 
can be calculated without resorting
to numerical differentiation.

Tables\, \ref{tab:bendinglevs}, \ref{tab:stretchlevs}, \ref{tab:comblevs-1}, and \ref{tab:comblevs-2}
compare some of the converged vibrational energy levels as
obtained by the gateway methods with the results of 
highly sophisticated DVR calculations by Bramley and Carrington \cite{bramley-carrington}. They  
employed Radau coordinates along with the same geometry and potential data as the ones used here.
The agreement between the gateway methods and Ref.\, \onlinecite{bramley-carrington} is very good.
The agreement between the different gateway methods is very good either. Vibrational levels
were also calculated without pseudo-potential. One can see that the energy levels 
involving only stretching are less influenced by removing or including the pseudo-potential than energy levels 
involving bending of the molecule. The contribution of the
pseudo-potential is more significant in the case of the Eckart Hamiltonian calculated by the projection method. 

By examining Figures\, \ref{fig:Gr1r1-r1}, \ref{fig:Gr1r1-phi}, \ref{fig:Gphiphi-r1}, \ref{fig:Gphiphi-phi}, and  
\ref{fig:bendpes} one might wonder how it is that different curves correspond to the same coordinates. Recall that
these coordinates are those of the configuration $\bm{a}$ but the function values correspond to those taken at 
Eckart configurations
$\bm{a}^{\textrm{E}}$. The Eckart coordinates are obtained by rotation,
$\bm{a}^{\textrm{E}}=\mathbb{U}\bm{a}$,
and  projection, $\bm{a}^{\textrm{E}}=\bm{P}\bm{a}$, respectively.
$\mathbb{U}$ is a $3N\times 3N$ block diagonal matrix
whose diagonal blocks are identical $3\times 3$ rotational matrices $\bm{U}(s)$ determined by the method of
Ref.\, \onlinecite{krasnos}. 
Rotation does not change bond angles and bond lengths whereas projection does. 
Therefore, the values of internal coordinates corresponding to the 
different configurations are related  as
\begin{gather}
s=s\left(\bm{a}(s)\right)=s\left(\mathbb{U}\bm{a}(s)\right)\neq \mathfrak{s}=s\left(\bm{P}\bm{a}(s)\right)
\end{gather}
Thus, for instance, one has the relationships
\begin{gather} 
V\left( s \right)=V\left(\bm{a}(s)\right)=V\left( \mathbb{U}\bm{a}(s) \right)
\neq V\left( \mathfrak{s} \right) =V\left( \bm{P}\bm{a}(s) \right). 
\end{gather}
Clearly, the Eckart Hamilton operators obtained by projection and rotation are 
different. Nevertheless, they must have the same spectrum, as, in fact, 
suggested by the results of numerical calculations.

To show that analytically, we start by noting that with $\bm{P}\bm{a}(s)$ 
being Eckart coordinates corresponding to a configuration with internal
coordinate values $\mathfrak{s}$, one can always find rotation $\mathbb{R}(\mathfrak{s})$ 
such that
$\bm{P}\bm{a}(s)=\mathbb{R}\bm{a}(\mathfrak{s})$, where 
$\mathbb{R}(\mathfrak{s})$ is a $3N\times 3N$ block diagonal matrix
whose diagonal blocks are identical $3\times 3$ rotational matrices $\bm{R}(\mathfrak{s})$ determined by the method of
Ref.\, \onlinecite{krasnos}.

The Eckart Hamilton operators 
$\hat{H}^{\textrm{E}}\left( \mathbb{U}\bm{a}(s); s, \hat{p}_{s}\right)$  and
$\hat{H}^{\textrm{E}}\left(\mathbb{R}\bm{a}(\mathfrak{s}); \mathfrak{s},\hat{p}_{\mathfrak{s}}\right)$
one obtains by using the Eckart coordinates 
$\bm{a}^{\textrm{E}}(s)$ and $\bm{a}^{\textrm{E}}(\mathfrak{s})$, respectively, 
are identical since they differ only in notation.  
Since rotation of the molecule in the coordinate system of axes 
cannot change the spectrum of a ro-vibrational Hamilton operator, it is also true that the spectrum of the 
non-Eckart Hamilton operator $\hat{H}\left(\bm{a}(s);  s,\hat{p}_{s}\right)$ is the same as that of the operator 
$\hat{H}^{\textrm{E}}\left(\mathbb{R}\bm{a}(\mathfrak{s});\mathfrak{s},\hat{p}_{\mathfrak{s}}\right)$.
There remains to show that the spectrum of 
$\hat{H}^{\textrm{E}}\left(\mathbb{R}\bm{a}(\mathfrak{s});\mathfrak{s},\hat{p}_{\mathfrak{s}}\right)$ is, the same as that of the 
operator 
$\hat{H}^{\textrm{E}}\left( \bm{P}\bm{a}(s); s, \hat{p}_{s}\right)$. If it can be shown 
that these operators are related by invertible 
coordinate
transformation, than, as a consequence they must have the same spectrum. A tentative proof is given below.

Let $ \mathscr{D}_{s} \subseteq
\mathscr{R}_{1}\times \mathscr{R}_{2}\times\mathscr{R}_{3N-6}$,
a subset of $3N-6$--tuples of real numbers $\mathscr{R}$,
be the domain accessible 
by the internal motions $s$. Projection of configurations $\bm{a}(s)$ implies the map
$\varphi  \left( \mathscr{D}_{s} \right) = \mathscr{D}_{\mathfrak{s}}$. 
Since in our case $\mathscr{D}_{\mathfrak{s}}$ is defined as the range of the map, the map is evidently onto. 
It is also one to one, if 
for every $\mathfrak{s} \in \mathscr{D}_{\mathfrak{s}}$ there is a unique
$s \in \mathscr{D}_{s}$ such that $ \mathfrak{s} = \varphi \left( s\right)$, that is $ \varphi \left( s \right)=
\varphi \left( s^{'} \right)$ implies $s=s^{'}$. 
Assume that $\varphi$ maps $s$ and $s^{'} (\neq s)$ into $\mathfrak{s}$, that is
$\bm{P}\bm{a}(s)=  \mathbb{R}(\mathfrak{s})\bm{a}\left( \mathfrak{s} \right)$ and
$\bm{P}\bm{a}(s^{'})=  \mathbb{R}^{'}(\mathfrak{s})\bm{a}\left( \mathfrak{s} \right)$.
By construction $\bm{P}\bm{a}(s)$ and $\bm{P}\bm{a}(s^{'})$ are Eckart coordinates.
Since the matrix  transforming $ \bm{a}\left( \mathfrak{s} \right)$ into
Eckart coordinates  is uniquely determined by the method of
Ref.\, \onlinecite{krasnos}
the matrices $\mathbb{R}(\mathfrak{s})$
and $\mathbb{R}^{'}(\mathfrak{s})$ must be equal to this matrix. Then $\bm{P}\bm{a}(s)= \bm{P}\bm{a}(s^{'})$ follows and
we have the relationships
\begin{gather}
\bm{a}(s)= \bm{Q}\bm{a}(s)+\bm{P}\bm{a}(s) 
\end{gather}
and
\begin{gather}
\bm{a}(s^{'})= \bm{Q}\bm{a}(s^{'})+\bm{P}\bm{a}(s),   
\end{gather}
where $\bm{Q}$ is the projection matrix  onto the translational-rotational 
subspace of the configuration space (i.e. $\bm{I}=\bm{P}+\bm{Q}$ with $\bm{I}$
denoting a $3N$ by $3N$ identity matrix).   
These equations show that changing the values of the internal coordinates from $s$ to $s^{'}$ causes displacements 
having no components in the vibrational space. But this cannot happen since $s$ are genuine internal coordinates whose
change should lead to non-zero displacement in the vibrational space as well. 
That is, $s^{'}$ must be equal to $s$. Therefore, the map $\mathfrak{s}=\varphi (s)$ is one to one and onto, 
which means that it is invertible\cite{szekeres}.

In a variational calculation any Hamiltonian, $\hat{H}$ or $\hat{H}^{\textrm{E}}$, with exact KEO
can be used. Nevertheless, the calculations may 
converge faster or more slowly depending on which Hamiltonian is employed. 
When one attempts to simplify a Hamiltonian 
by introducing approximations the quality of the approximate Hamiltonian derived may strongly depends on to which Hamiltonian
are the approximations invoked. Eckart Hamiltonians may have advantage, whether the rate of convergence or the quality of
the derived approximate Hamiltonians are considered. 
However, there are many different Eckart Hamiltonians (of which two have been considered in the
numerical examples).
Therefore, it may be useful to find a way of selecting a unique Eckart Hamiltonian out of 
the many different ones. This question can be reduced to finding Eckart displacements optimal in some appropriate sense.
It is addressed in the next Section. 

\begin{table*}[!htb]
\begin{threeparttable}
\caption{Bending energy levels} \label{tab:bendinglevs} 
\begin{ruledtabular}
\begin{tabular}{lccccc}
( n$_{1}$ n$_{2}$ n$_{3}$ )   &  BC   & Gateway rot  & {\textbf {\em Gateway proj}} & Gateway rot, no $V_{\textrm{ps}}$ 
& {\textbf {\em Gateway proj, no}} $V_{\textrm{ps}}$ \\ 
\hline 
(0 0 0) & 4630.3465   & 4630.295443  & {\textbf {\em 4630.295446}}   &  4650.23     
&  {\textbf {\em 4638.90}}  \\
        &           & $\Delta $         & $\Delta $             & $\Delta $ &   $\Delta $  \\
(0 1 0) & 1594.32   & 0.02  & {\textbf {\em 0.02}}      &  -0.78     
&  {\textbf {\em -2.24}}\\
(0 2 0) & 3152.01   & 0.04  & {\textbf {\em 0.03 }}     & -1.85    
&  {\textbf {\em -4.58 }}\\
(0 3 0) & 4667.70   & 0.06 & {\textbf {\em 0.05}}     &  -3.4      
&  {\textbf {\em -7.27}} \\
(0 4 0) & 6134.11   & 0.78  & {\textbf {\em 0.07} }     &  -5.19      
&  {\textbf   {\em-10.85}}\\
(0 5 0) & 7539.79   & 0.06  & {\textbf {\em 0.01} } &  -10.18     
&  {\textbf  {\em -16.84}} \\
Basis size &  & $25\times 41\times 41$ &  $23\times 51\times 51$ & $25\times 41\times 41$ &  $23\times 51\times 51$ 
\end{tabular}
\end{ruledtabular}
\begin{tablenotes}
      \small
      \item a) $n_{1},n_{2}$ and $n_{3}$ are the number of quanta in the symmetric stretching, the bending, 
                   and the asymmetric stretching vibrations, respectively.\\
      
      \item b) Column BC contains energy values from Ref.\, \onlinecite{bramley-carrington}. 
        The assignment is from Ref.\, \onlinecite{fernley-et-al}. 
\\
      
      \item c) $\Delta$ denotes the difference between the energy values obtained by BC and the gateway methods. \\
        
      \item d) Energy levels are given with respect to the gound state energy in units of wave numbers (cm$^{-1}$).\\

      \item e) When the rotation and the projection methods were employed 
              the volumes of internal coordinate space sampled by the grid had been 
               $ \phi \in \left( 51^{\circ},~160.4^{\circ}\right),~ r_{1},r_{2} \in \left(0.6\text{\AA},~2.535\text{\AA}\right)$ 
and
              $ \phi \in \left( 54^{\circ},~155^{\circ}\right),~ r_{1},r_{2} \in \left(0.6\text{\AA},~2.535\text{\AA}\right)$, 
respectively. \\

            \item f) Basis size: $N_{\phi}\times N_{r_{1}}\times  N_{r_{2}}$, where $N_{\phi}$ and $N_{r_{1}}$,
            $~N_{r_{2}}$ are the number of basis functions for the bending and  stretching vibrations.  

    \end{tablenotes}
\end{threeparttable}
\end{table*}

\begin{table*}[!htb]
\begin{threeparttable}
\caption{Stretching energy levels} \label{tab:stretchlevs} 
\begin{ruledtabular}
\begin{tabular}{lccccc}
( n$_{1}$ n$_{2}$ n$_{3}$ )        &  BC   & Gateway rot & {\textbf {\em Gateway proj}} & Gateway rot, no $V_{\textrm{ps}}$ & 
{\textbf {\em Gateway proj, no}} $V_{\textrm{ps}}$\\
\hline 
        &           & $\Delta$  &  $\Delta$    & $\Delta $    &   $\Delta $  \\
(1 0 0) & 3656.49   &   0.04   & {\textbf {\em 0.04}} &   0.43  &   
{\textbf  {\em 0.21}} \\
(2 0 0) & 7202.67   &   0.08  & {\textbf {\em 0.08}}  &  0.86   &    
{\textbf  {\em 0.44}}\\
(3 0 0) & 10602.76   &  0.1  & {\textbf {\em 0.1}} &   1.3   &   
{\textbf  {\em 0.7}}  \\
(4 0 0) & 13829.70   &   0.14  & {\textbf  {\em 0.14}} &  -0.56   &   
{\textbf  {\em 0.91}}\\
(5 0 0) & 16899.45   &   0.61 & {\textbf  {\em 0.01}} &   1.95  &   
{\textbf  {\em -0.22}} \\
(0 0 1) & 3755.92   &   0.04  & {\textbf {\em 0.04} } &  0.51  &   
{\textbf  {\em 0.38}}\\
(0 0 2) & 7444.93   &   0.08   & {\textbf  {\em 0.08} } &    -0.01  &    
{\textbf  {\em 0.7}}\\
(0 0 3) & 11034.09   &   0.12  & {\textbf  {\em 0.12 }} & 1.47   &    
{\textbf {\em 1.02}}\\
(0 0 4) & 14541.30   &   0.14  & {\textbf {\em 0.15 }} &   -2.57   &    
{\textbf {\em -0.42}}\\
(0 0 5) & 17954.91   &    0.18  & {\textbf  {\em 0.18} } &   2.36  &    
{\textbf  {\em 1.53}} 
\end{tabular}
\end{ruledtabular}
\end{threeparttable}
\end{table*}

\begin{table*}[!htb]
\begin{threeparttable}
\caption{Combinations} \label{tab:comblevs-1}
\begin{ruledtabular}
\begin{tabular}{lccccc}
( n$_{1}$ n$_{2}$ n$_{3}$ )  &  BC  & Gateway rot & {\textbf {\em Gateway proj}} & Gateway rot, no $V_{\textrm{ps}}$ & 
{\textbf  {\em Gateway proj, no}} $V_{\textrm{ps}}$\\
\hline 
        &           & $\Delta $        & $\Delta $       &  $\Delta $  &   $\Delta $  \\
(1 1 0) & 5234.29   &   0.06   & {\textbf  {\em 0.06} }  &   -0.33   &  {\textbf {\em -1.97}} \\
(1 2 0) & 6775.03   &   0.1   & {\textbf {\em 0.08} }  &   -1.33  &   {\textbf {\em -4.23}} \\
(1 3 0) & 8273.24   &   0.4 & {\textbf  {\em 0.07} }  &  -2.59  &   {\textbf {\em -6.91}}\\
(1 4 0) & 9719.75   &  1.47   & {\textbf {\em -0.4} }  &  -4.15  &  {\textbf {\em  -11.54}} \\
(0 1 1) & 5332.06   &  0.06  & {\textbf {\em  0.06}}  &  -0.19   &    {\textbf   {\em -1.75}}\\
(0 1 2) & 9002.14   &  0.1 & {\textbf {\em  0.1} }  &  0.35  &    {\textbf {\em  -1.35}}\\
(0 1 4) & 16057.58   &  0.17  & {\textbf {\em 0.18 }}  & 1.38 &    {\textbf {\em -0.61}}\\
(0 1 5) & 19449.25   &  0.2  & {\textbf {\em 0.19} }  & -0.13  &   {\textbf  {\em -0.28}} \\
(0 6 1) & 12571.35   &  0.2   & {\textbf {\em 0.22}}  & 0.74    &   {\textbf {\em   -0.88}}
\end{tabular}
\end{ruledtabular}
\end{threeparttable}
\end{table*}

\begin{table*}[!htb]
\begin{threeparttable}
\caption{Combinations} \label{tab:comblevs-2}
\begin{ruledtabular}
\begin{tabular}{lccccc}
( n$_{1}$ n$_{2}$ n$_{3}$ )        &  BC  & Gateway rot & {\textbf {\em  Gateway proj}} & Gateway rot, no $V_{\textrm{ps}}$ & 
{\textbf {\em Gateway proj, no}} $V_{\textrm{ps}}$\\
\hline 
        &           & $\Delta $        & $\Delta $      & $\Delta $        &   $\Delta $  \\
(1 1 1) & 8809.59   &   0.09   & {\textbf {\em 0.09 }}  & 0.23    &     {\textbf {\em -1.53}}\\
(1 2 1) & 10332.40   &  0.12  & {\textbf  {\em 0.11 }} & -0.65   &    {\textbf {\em -3.61 }}\\
(1 3 1) & 11815.47   &  0.34  & {\textbf {\em 0.26} } & -1.76  &     {\textbf {\em  -5.86}}\\
(2 1 1) & 12156.52   &  0.13  & {\textbf  {\em 0.12} } & 0.67  &    {\textbf {\em   -1.3}} \\
(3 1 1) & 15355.27   &  0.16    & {\textbf {\em  0.15} } & 1.04  &     {\textbf {\em -1.18}}\\
(1 1 2) & 12408.42   &  -0.21   & {\textbf  {\em -0.4} }  & -0.09   &  {\textbf {\em  -1.09}}\\
(1 2 2) & 13911.72   &  0.20  & {\textbf  {\em -0.47} } & -0.16  &     {\textbf {\em  -3.27}}\\
(2 2 2) & 17226.08   &  -0.08  & {\textbf {\em  0.79} }& -0.16  &    {\textbf  {\em -3.7}} \\
(1 1 3) & 15839.10   &  0.21  & {\textbf  {\em -0.24} } & 1.22  &    {\textbf {\em  -0.9}} 
\end{tabular}
\end{ruledtabular}
\end{threeparttable}
\end{table*}

\section{Geometry of Eckart conditions: Optimal Eckart displacements}

\begin{figure*}[!htb]
\begin{center}
\includegraphics[scale=0.6,keepaspectratio]{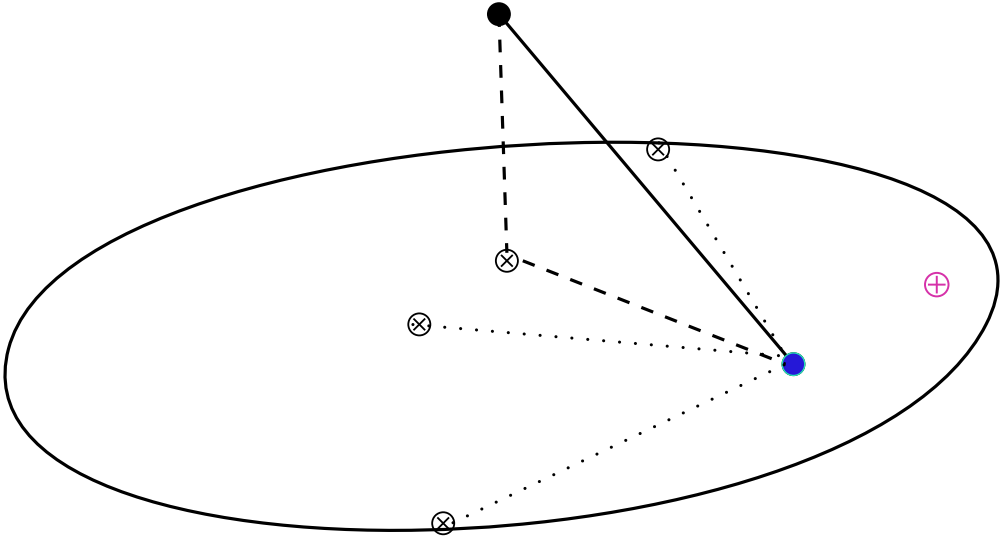}
\caption{ The ellipse represents the vibrational space, a subspace in the translational reduced 
configuration space,  defined by the Eckart conditions.
Since they obey the Eckart conditions, the origin of the MS,
${\textit vilolet}\oplus$, as well as the point 
${\textit blue}{\bullet} \equiv \bm{m}^{1/2}\bm{a}^{0}$ corresponding to the reference configuration are in
this subspace. A general point of configuration space is $\bullet \equiv \bm{m}^{1/2} \bm{a}$. To find Eckart coordinates
one must find a map $\mathcal{M}$ such that $\otimes = \mathcal{M} \bullet$.}
\label{fig:geomofeckart-1}
\end{center}
\end{figure*}

\begin{figure*}[!htb]
\begin{center}
\includegraphics[scale=0.6,keepaspectratio]{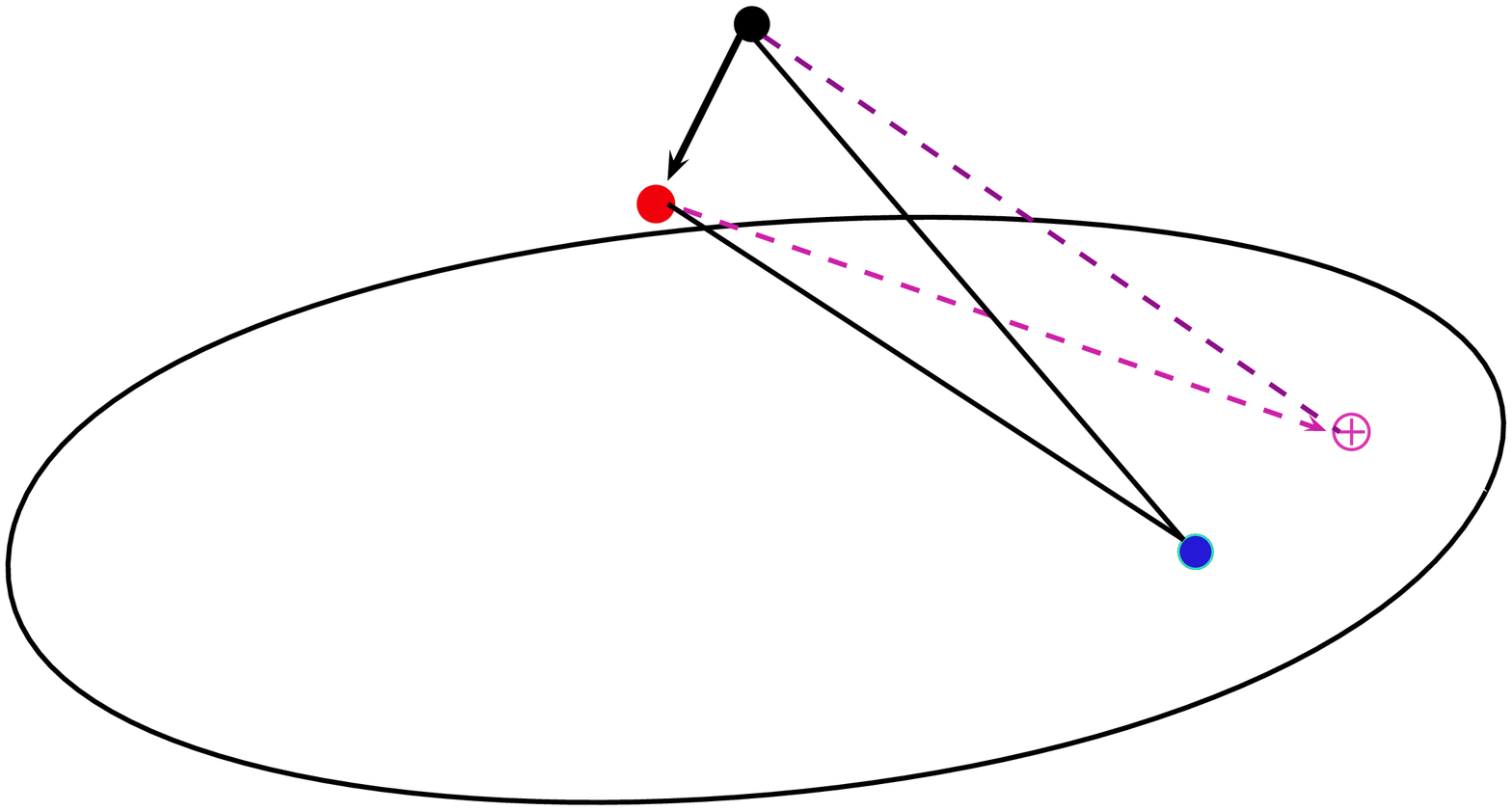}
\caption{Rotation map. ${\textit red}\bullet \equiv \mathbb{U}\bm{m}^{1/2}\bm{a}~$,
$\bm{m}^{1/2}\bm{d} = {\textit red}\bullet - \textit{blue}\bullet = \bm{U}\bm{m}^{1/2}\bm{a}-\bm{m}^{1/2}\bm{a}^{0}~$,
$\mathrm{MWSD}  =  \lVert {\textit red}\bullet - {\textit blue}\bullet \rVert^{2} =\left[ \bm{d} \right]^{T}\bm{m}\bm{d}~$
} \label{fig:geomofeckart-2}
\end{center}
\end{figure*}

Pictures are often helpful in explaining and learning ideas. Figure\, \ref{fig:geomofeckart-1} gives a simplified pictorial
representation of the geometry of Eckart conditions. One can see immediately that to obtain Eckart coordinates (and displacements)
one must find a map $\mathcal{M}$ mapping a general point of  the translation reduced configuration space, 
$\bullet$, into a point, $\otimes$, of the vibrational space.    
One such map, namely, $\mathcal{M}=\bm{P}$, has been already considered in the previous Sections.  
Note that $\bm{P}$ solves the optimization problem
$\min\limits_{\mathcal{M}} \lVert \bullet - \otimes \rVert^{2}$, and to each  point $\bullet$ there corresponds a unique point
$\otimes$ in vibrational space.  ( $\lVert v \lVert$ denotes the length of a vector $v$.)
Now imagine connecting the point $\oplus$ representing the origin of the MS  with the 
point $\bullet$ by a straight line and drawing the hypersphere of radius $\lVert \bullet-\oplus \lVert$ 
in the (translation reduced) configuration space.
Intersections of this sphere with the vibrational space give Eckart coordinates. In general, however, this  does not give a
unique image of the point $\bullet$. Therefore it is not a map \cite{schutz}. To simplify, let us restrict to transformations of the form
$ \mathbb{U}= \mathrm{diag} \left( \bm{U}_{1},\bm{U}_{2},\ldots, \bm{U}_{3N-6}\right)$
with $\bm{U}_{1}=\bm{U}_{2}=\cdots=\bm{U}_{3N-6}=\bm{U}$ and $\bm{U} \in SO(3)$.      
Figure\, \ref{fig:geomofeckart-2} shows such a transformation ${\textit red}\bullet = \mathbb{U}\bullet$ corresponding 
to mass weighted displacement 
$\bm{m}^{1/2}\bm{d} = {\textit red}\bullet - 
{\textit blue}\bullet = \mathbb{U}\bm{m}^{1/2}\bm{a}-\bm{m}^{1/2}\bm{a}^{0}$. 
The square of the length of this displacement, which may be called mass weighted squared displacement (MWSD), is 
$\mathrm{MWSD}  =  \lVert {\textit red}\bullet - {\textit blue}\bullet \rVert^{2} =\left[ \bm{d} \right]^{T}\bm{m}\bm{d}$.  
It has been shown in Ref.\, \onlinecite{krasnos}  that the displacement vector calculated by $\mathbb{U}$ 
solving the problem of minimization $\min\limits_{\mathbb{U}} \left( \left[ \bm{d} \right]^{T}\bm{m}\bm{d}\right)$
obeys the Eckart
conditions. In other words, of all equivalent configurations,
that is among all $\mathbb{U}\bm{a}$, the closest to the reference
configuration is related to the reference configuration by
Eckart displacements. Experience shows that it is uniquely determined.

\begin{figure*}[!htb]
\begin{center}
\includegraphics[scale=0.6,keepaspectratio]{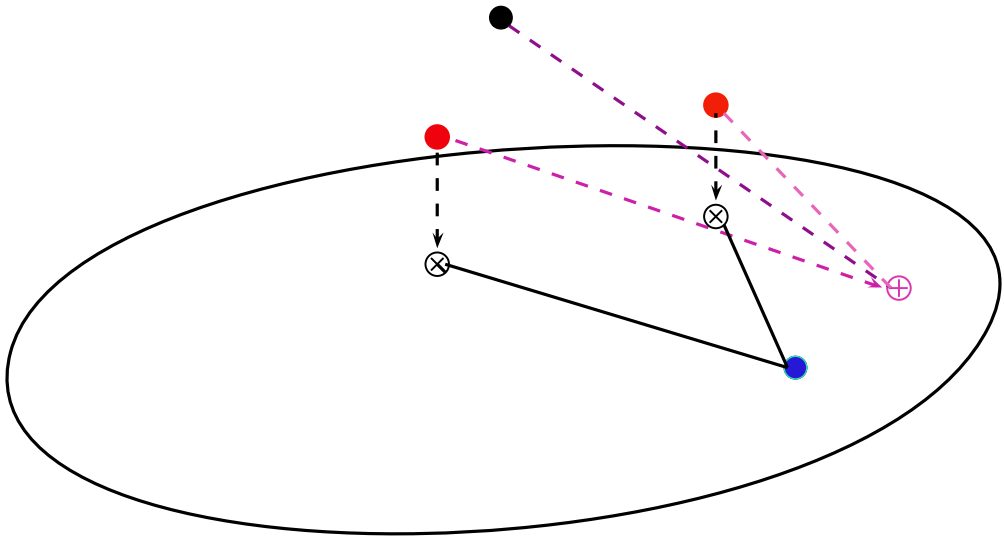}
\caption{Combination of projection and rotation maps.} \label{fig:geomofeckart-3}
\end{center}
\end{figure*}

Projection and rotation may be combined to generate  Eckart coordinates as depicted in Figure\, \ref{fig:geomofeckart-3}.
Observe that
the points 
${\textit red} \otimes  = \bm{P}\mathbb{U} \bm{m}^{1/2}\bm{a} $
correspond to Eckart configurations for any choice of $\mathbb{U}$.

Therefore one can ask for an $\mathbb{U}$ leading to
an Eckart configuration which is the closest to the
reference configuration among all Eckart configurations which
one can derive from a single distorted configuration by rotation. 
Translated into mathematical form it means
minimization (with respect to $\mathbb{U}$) of the quantity
\begin{eqnarray}
{\textit blue} {\mathrm{MWSD}} & = & \left[ \bm{d}^{\mathrm{E}} \right]^{T}\bm{m}\bm{d}^{\mathrm{E}}
\nonumber \\
& = &
 \sum_{j=1}^{3N-6} \lvert c_{j} \rvert^{2} \nonumber \\
& = & \left[ \bm{m}^{1/2} \left( \mathbb{U}\bm{a}-\bm{a}^{0}\right)\right]^{T}\bm{P}
\bm{m}^{1/2} \left( \mathbb{U}\bm{a}-\bm{a}^{0}\right) .
\end{eqnarray}
Since, as explained in Ref.\, \onlinecite{szalay-eckart}, the projection matrix onto the vibrational space can be
also expressed as
\begin{gather}
\bm{P} = \bm{1}-{\bar {\cal T}}_{\alpha}^{T}{\bar {\cal T}}_{\alpha} - {\bar {\cal R}}_{\alpha}^{T}{\bar {\cal R}}_{\alpha},
\end{gather}
where
\begin{gather}
{\bar {\cal T}}_{\alpha,\gamma n} = M^{-1/2} m_{n}^{1/2}
\delta_{\alpha \gamma}
\end{gather}
with $M$ denoting the mass of the molecule, and
\begin{gather}
{\bar {\cal R}}_{\alpha, \gamma n} = \left[ {\mathbf{I}^{0}}^{-1/2} \right]_{\alpha\delta}
\varepsilon_{\delta \beta \gamma } m_{n}^{1/2} a^{0}_{\beta n},
\end{gather}
where $\varepsilon_{\delta \beta \gamma }$ is the Levi-Civita tensor and 
$\mathbf{I}^{0}$ stands for the rotational tensor of inertia of the reference configuration, 
one can obtain that
\begin{widetext}
\begin{gather}
 {\textit blue} {\mathrm{MWSD}}= 
\lVert \bm{m}^{1/2} \left( \mathbb{U}\bm{a}-\bm{a}^{0}\right)
\rVert^{2}
- \lVert 
{\bar {\cal R}}_{\alpha}
\bm{m}^{1/2}\mathbb{U}\bm{a} \rVert^{2} \nonumber \\
=
\sum_{n=1}^{N} m_{n} \lVert  \bm{U}\bm{a}_{n}-\bm{a}^{0}_{n} \rVert^{2}
-
\sum_{p=1}^{N} m_{p} \left( \bm{a}^{0}_{p}\times \bm{U}\bm{a}_{p} \right)_{\alpha}
\left[ {\mathbf{I}^{0}}^{-1} \right]_{\alpha\beta} \sum_{n=1}^{N} m_{n}\left( \bm{a}^{0}_{n}\times
 \bm{U}\bm{a}_{n} \right)_{\beta}. \label{form1}
\end{gather}
\end{widetext}
In deriving Eq.\, (\ref{form1}) 
use has been made of the fact the $\bm{a}^{0}_{n}$ and
$\bm{a}_{n}$ obey the Eckart conditions and the translational Eckart conditions, respectively.

\begin{figure*}[!htb]
\begin{center}
\includegraphics[angle=270,scale=0.5,keepaspectratio]{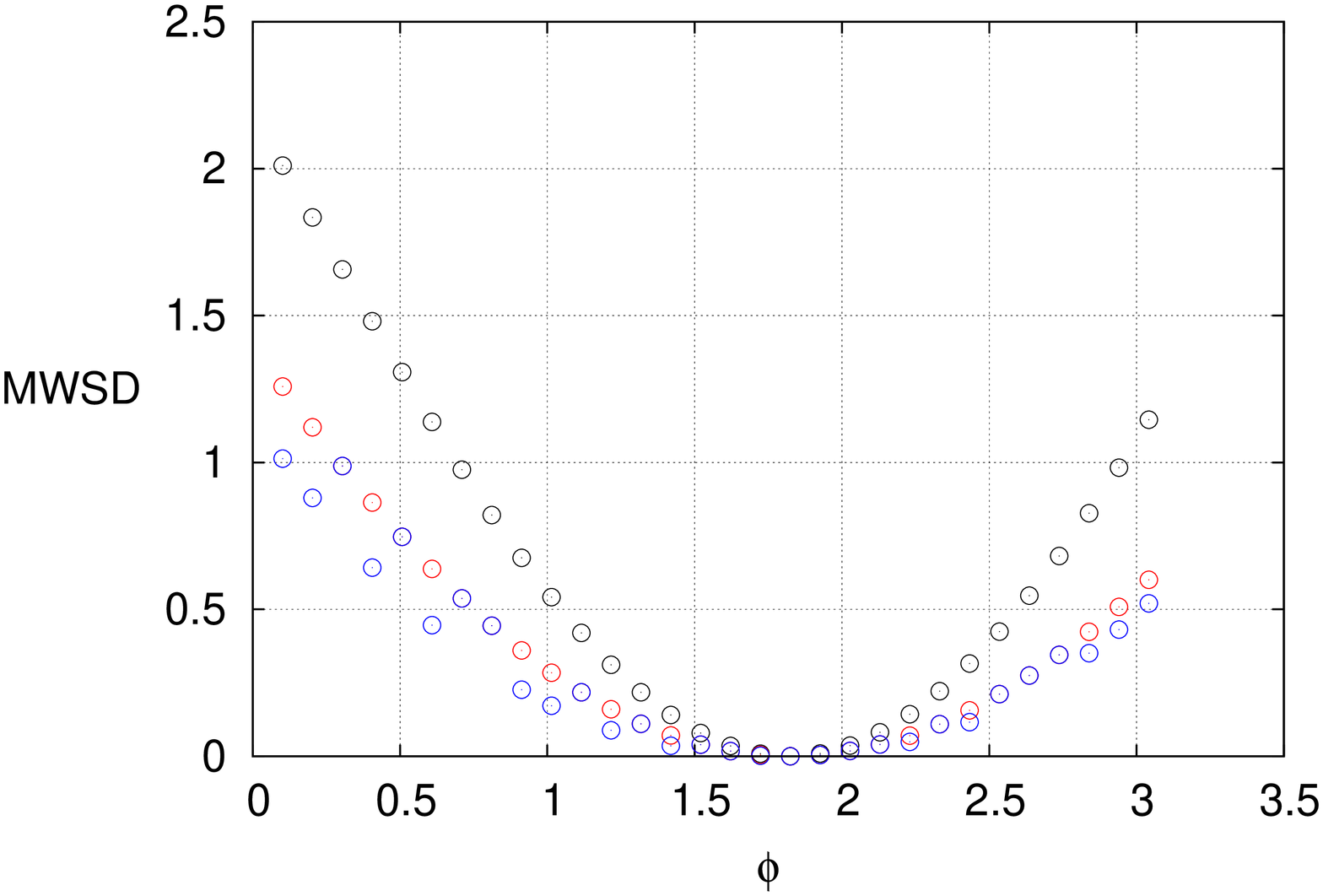}
\caption{ The MWSD for H$_{2}$O as function of the bending angle when the bond lengths are fixed at their equilibrium
values.  
$\color{black}{\odot}  =  \mathrm{MWSD}~$,  
${\textit red}{\odot} =  \min\limits_{\bm{U}} \mathrm{MWSD}~$, 
${\textit blue}{\odot} =  \min\limits_{\bm{U}} {\textit blue}{\mathrm{MWSD}}.$
} \label{fig:mwsd}
\end{center}
\end{figure*}

\begin{figure*}[!htb]
\begin{center}
\includegraphics[angle=270,scale=0.5,keepaspectratio]{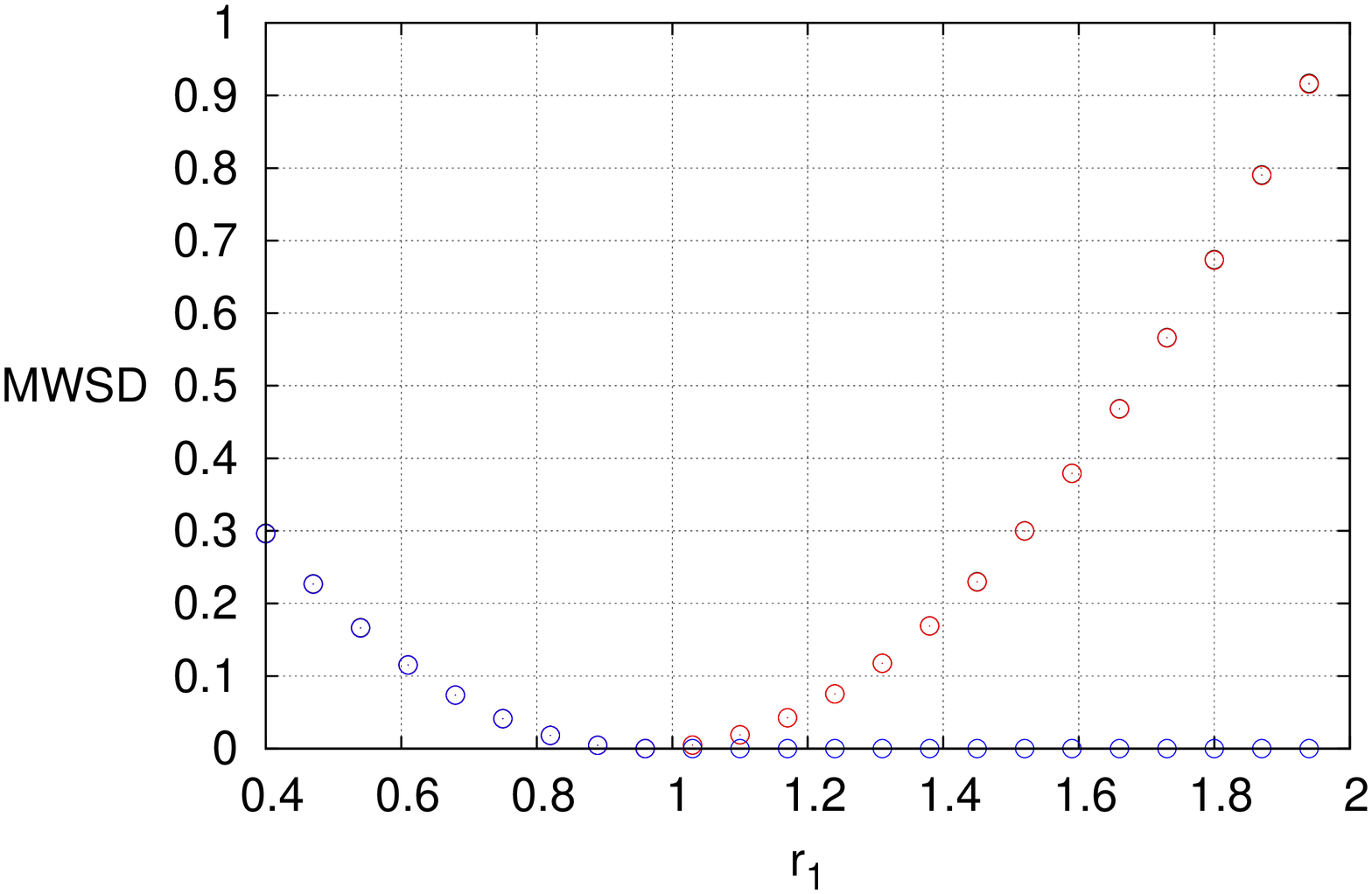}
\caption{
The MWSD for H$_{2}$O as function of the bond length $r_{1}$ when the bond length $r_{2}$ and the bending angle are
fixed at their equilibrium
values.
$\color{black}{\odot}  =  \mathrm{MWSD}~$,
${\textit red}{\odot} =  \min\limits_{\bm{U}} \mathrm{MWSD}~$,
${\textit blue} {\odot} =  \min\limits_{\bm{U}} {\textit blue}{\mathrm{MWSD}}.$
(The black and red curves coincide.)
} \label{fig:MWSD-r1}
\end{center}
\end{figure*}

If $\bm{U}$ is chosen such that $\bm{U}\bm{a}_{n}$ are Eckart
coordinates, then Eq.\, (\ref{form1}) shows that ${\textit blue} {\mathrm{MWSD}} =\min\limits_{\bm{U}} \mathrm{MWSD} $. Therefore, it follows 
that 
\begin{gather}
\min\limits_{\bm{U}}{\textit blue} {\mathrm{MWSD}} \le \min\limits_{\bm{U}} \mathrm{MWSD}.\label{ineq}
\end{gather}
Results of numerical calculations shown in Figures\, \ref{fig:mwsd} and 
\ref{fig:MWSD-r1} give numerical evidence. Some details of solving the optimization problem 
$\min\limits_{\bm{U}}{\textit blue}{\mathrm{MWSD}}$
are described in 
Appendix \ref{iterativemin}. One can see in Figure\, \ref{fig:mwsd} that the optimization occasionally bogs down at local
minima. At these points the optimization must be restarted by new initial parameters and the global minimum can be found.
Nevertheless, the numerical results presented do confirm the inequality Eq.\, (\ref{ineq}). 

\section{Summary}
The Eckart conditions treated as what they are,
a homogeneous system of linear equations, has led to the gateway Hamiltonian method
\cite{szalay-gateway,szalay-aspects}.
In this approach the conditions define the space of vibrations
and one uses them to determine the molecule fixed system coordinates of the atoms as functions of vibrational coordinates,
and, eventually, to derive Eckart ro-vibrational Hamiltonians. 
The derivations 
assume that a basis for the vibrational space is already known. A prerequisite for practical use of the
gateway method is, therefore, the availability of such a basis set. 
In the present work general analytical formulas of such a basis set
have been derived and some of the practical advantages of the gateway Hamiltonian method have been numerically
demonstrated. It has been pointed out that 
there is an infinite number of Eckart KEOs corresponding to any given set of curvilinear internal coordinates. 
It is suggested that one should use 
the KEO corresponding to in some sense optimally defined Eckart displacements. 
A possible definition of optimal Eckart displacements has been introduced and illustrated with numerical examples.

The results described may give food for thoughts:
\begin{itemize} 
\item Construction and application of a Hamilton operator with optimal Eckart displacements. 
\item The geometrical interpretation of the Eckart conditions may elucidate and can help 
deriving the relationships of Eckart coordinates corresponding either to different
electronic states \cite{duschinsky} or isotopic species\cite{meier}  of a molecule. 
\item Conditions other than the Eckart, e.g.
those for using the instantaneous principal axis system, might be treated 
similarly by replacing linear algebra with the appropriate mathematical technique(s).
\end{itemize}

\begin{acknowledgments}
The author is indebted to Ildik\'o Horv\'ath for her support. 
\end{acknowledgments}

\appendix

\section{Analytical solution of the Eckart conditions} \label{eckart-sol}

Let $\bm{t}^{\alpha}$ and $\bm{r}^{\alpha}$ denote the row vectors of elements
\begin{gather}
t^{\alpha}_{\beta n} = m_{n}\delta_{\alpha\beta},
\end{gather}
and
\begin{gather}
r^{\alpha}_{\beta n} = \epsilon_{\alpha\gamma\beta} m_{n} a^{0}_{\gamma n}.
\end{gather}

By employing this notation the Eckart conditions,
\begin{subequations}
\label{eckartequations}
\begin{gather}
\sum_{n=1}^{N} m_{n} d_{xn} = 0, \label{tx}\\
\sum_{n=1}^{N} m_{n} d_{xn} = 0, \label{ty}\\
\sum_{n=1}^{N} m_{n} d_{zn} = 0, \label{tz}\\
\sum_{n=1}^{N} \left( a^{0}_{yn}m_{n}d_{zn}-m_{n}d_{yn}a^{0}_{zn} \right) = 0, \label{rx}\\
\sum_{n=1}^{N} \left( a^{0}_{xn}m_{n}d_{zn}-m_{n}d_{xn}a^{0}_{zn} \right) = 0, \label{ry}\\
\sum_{n=1}^{N} \left( a^{0}_{xn}m_{n}d_{yn}-m_{n}d_{xn}a^{0}_{yn}  \right) = 0, \label{rz}
\end{gather}
\end{subequations}
 can be written as a system of linear equations
\begin{gather}
\bm{E}\bm{d} = 0, \label{linsys}
\end{gather}
with
\begin{widetext}
{\small
\begin{gather}
\bm{E} = \left( 
\begin{array}{c}
\bm{t}^{x} \\
\bm{t}^{y} \\
\bm{t}^{z} \\
\bm{r}^{x} \\ 
\bm{r}^{y} \\
\bm{r}^{z}
\end{array}
\right) 
\nonumber \\
=\left(
\begin{array}{cccccccccccc}
m_{1} & 0 & 0 & m_{2} & 0 & 0 & . & . & .  & m_{N} & 0 & 0\\
0     & m_{1} & 0 & 0     & m_{2} & 0 & . & . & . & 0 & m_{N} & 0   \\
0 & 0 & m_{1} & 0 & 0 & m_{2} & . & . & .  & 0 & 0 & m_{N} \\
0 & -m_{1}a_{z1}^{0} & m_{1}a_{y1}^{0} & 0 & -m_{2}a_{z2}^{0} & m_{2}a_{y2}^{0}
& . & . & . & 0 & -m_{N}a_{zN}^{0} & m_{N}a_{yN}^{0} \\
m_{1}a_{z1}^{0}     & 0 & -m_{1}a_{x1}^{0} & m_{2}a_{z2}^{0}  & 0 & -m_{2}a_{x2}^{0}
& . & . & .   & m_{N}a_{zN}^{0} & 0 & -m_{N}a_{xN}^{0}\\
-m_{1}a_{y1}^{0} & m_{1}a_{x1}^{0} & 0 & -m_{2}a_{y2}^{0} & m_{2}a_{x2}^{0} & 0 & . & . & .
& -m_{N}a_{yN}^{0} & m_{N}a_{xN}^{0}  & 0
\end{array}
\right).
\end{gather} 
}
\end{widetext}
Since Eq.\, (\ref{linsys}) is a homogeneous system, its general solution, $\bm{d}^{\textrm{E}}$, can be written as
\begin{gather}
\bm{d}^{\textrm{E}} = \sum_{j=1}^{K} b_{j}\bm{h}^{\textrm{E},j},
\end{gather}
where $K=3N-\textrm{rank} (\bm{E})$, $b_{j}$ are free variables, and $\bm{h}^{\textrm{E},j}$ are particular solutions 
(that is they obey the Eckart conditions). We shall determine $\bm{h}^{\textrm{E},j}$. Then, $\bm{h}^{\textrm{E},j}$ are
employed to calculating a basis orthonormal with mass weighting and spanning the vibrational space. 

Gaussian elimination \cite{carl-meyer} applied to Eq.\, (\ref{linsys}) gives 
\begin{gather}
\bm{C} = \left(
\begin{array}{c}
\bm{t}^{x} \\
\bm{t}^{y} \\
\bm{t}^{z} \\
\bm{r}^{y}-a_{z1}^{0}\bm{t}^{x}+a_{x1}^{0}\bm{t}^{z} \\ 
\bm{r}^{x}+a_{z1}^{0}\bm{t}^{y}-a_{y1}^{0}\bm{t}^{z} \\
\bm{r}^{z}+a_{y1}^{0}\bm{t}^{x}-a_{x1}^{0}\bm{t}^{y} -\frac{-a_{y2}^{0}+a_{y1}^{0}}{a_{z2}^{0}-a_{z1}^{0}}\bm{v}^{x}
+\frac{a_{x2}^{0}-a_{x1}^{0}}{a_{z2}^{0}-a_{z1}^{0}}\bm{v}^{y}
\end{array}
\right),
\end{gather}
where
\begin{subequations}
\label{vx}
\begin{gather}
\bm{v}^{x} = \bm{r}^{y}-a_{z1}^{0}\bm{t}^{x}+a_{x1}^{0}\bm{t}^{z}, \\
\textrm{with} \nonumber \\
v^{x}_{xn} = m_{n}a^{0}_{zn}-m_{n}a^{0}_{z1},  \\
v^{x}_{yn} = 0,  \\
v^{x}_{zn} = -m_{n}a^{0}_{xn}+m_{n}a^{0}_{x1},  
\end{gather}
\end{subequations}
and
\begin{subequations}
\label{vy}
\begin{gather}
\bm{v}^{y} = \bm{r}^{x}+a_{z1}^{0}\bm{t}^{y}-a_{y1}^{0}\bm{t}^{z}, \\
\textrm{with} \nonumber \\
v^{y}_{xn} = 0,  \\
v^{y}_{yn} = -m_{n}a^{0}_{zn}+m_{n}a^{0}_{z1},  \\
v^{y}_{zn} = m_{n}a^{0}_{yn}-m_{n}a^{0}_{y1},
\end{gather}
\end{subequations}

By introducing 
\begin{widetext}
\begin{subequations}
\label{vz}
\begin{gather}
\bm{v}^{z} = 
\bm{r}^{z}+a_{y1}^{0}\bm{t}^{x}-a_{x1}^{0}\bm{t}^{y} -\frac{-a_{y2}^{0}+a_{y1}^{0}}{a_{z2}^{0}-a_{z1}^{0}}\bm{v}^{x}
+\frac{a_{x2}^{0}-a_{x1}^{0}}{a_{z2}^{0}-a_{z1}^{0}}\bm{v}^{y}, \\
\textrm{that is} \nonumber \\
v^{z}_{xn} = -m_{n}a^{0}_{yn}+m_{n}a^{0}_{y1}-\frac{-a_{y2}^{0}+a_{y1}^{0}}{a_{z2}^{0}-a_{z1}^{0}}
\left( m_{n}a^{0}_{zn}-m_{n}a^{0}_{z1} \right),  \\
v^{z}_{yn} = m_{n}a^{0}_{xn}-m_{n}a_{x1}^{0}+\frac{a_{x2}^{0}-a_{x1}^{0}}{a_{z2}^{0}-a_{z1}^{0}}
\left( -m_{n}a^{0}_{zn}+m_{n}a^{0}_{z1} \right),  \\
v^{z}_{zn} =-\frac{-a_{y2}^{0}+a_{y1}^{0}}{a_{z2}^{0}-a_{z1}^{0}}\left( -m_{n}a^{0}_{xn}+m_{n}a^{0}_{x1}\right)
+\frac{a_{x2}^{0}-a_{x1}^{0}}{a_{z2}^{0}-a_{z1}^{0}}\left( m_{n}a^{0}_{yn}-m_{n}a^{0}_{y1} \right), 
\end{gather}
\end{subequations}
\end{widetext}
one can write $\bm{C}$ as
\begin{widetext}
{\small
\begin{gather}
\bm{C} = 
\left(
\begin{array}{cccccccccccccccccc} 
m_{1} & 0     & 0     & m_{2}         & 0          & 0          & m_{3}     & 0         & 0          &  m_{4} & 0 & 0 
& . & . & . &  m_{N} & 0 & 0 \\
0     & m_{1} & 0     & 0             & m_{2}      & 0          & 0         & m_{3}     & 0          & 0 &   m_{4} & 0 
& . & . & . & 0 &   m_{N} & 0 \\
0     & 0     & m_{1} & 0             & 0          & m_{2}      & 0         & 0         & m_{3}      & 0 & 0 &  m_{4} 
& . & . & . & 0 & 0 &  m_{N}\\
0     & 0     & 0     & v^{x}_{x2}    & 0          & v^{x}_{z2} & v^{x}_{x3}& 0 & v^{x}_{z3} & v^{x}_{x4} & 0  & 
v^{x}_{z4} & . & . & .  & v^{x}_{xN} & 0  & v^{x}_{zN} \\
0     & 0     & 0     & 0             & v^{y}_{y2} & v^{y}_{z2} & 0         & v^{y}_{y3}& v^{y}_{z3} & 0 & v^{y}_{y4} & 
v^{y}_{z4} & . & . & . & 0 & v^{y}_{yN} & v^{y}_{zN} \\
0     & 0     & 0     & 0             & 0          & 0          & v^{z}_{x3}& v^{z}_{y3}& v^{z}_{z3} & v^{z}_{x4} & v^{z}_{y4} & 
v^{z}_{z4} & . & . & . & v^{z}_{xN} & v^{z}_{yN} & v^{z}_{zN} 
\end{array}
\right).
\end{gather}
}
\end{widetext}

Thus, we have the system of equations
{\small
\begin{subequations}
\label{eqsys}
\begin{gather}
\sum_{n=1}^{N} m_{n} d_{xn} = 0, \\
\sum_{n=1}^{N} m_{n} d_{xn} = 0, \\
\sum_{n=1}^{N} m_{n} d_{zn} = 0, \\
\sum_{n=2}^{N} \left( v^{x}_{xn}d_{xn}+v^{x}_{zn}d_{zn} \right) = 0, \\
\sum_{n=2}^{N} \left( v^{y}_{yn}d_{yn}+v^{y}_{zn}d_{zn} \right) = 0, \\
\sum_{n=3}^{N} \left( v^{z}_{xn}d_{xn}+v^{z}_{yn}d_{yn}+v^{z}_{zn}d_{zn} \right) = 0.
\end{gather}
\end{subequations}
}

Then,
\begin{widetext}
{\small
\begin{subequations}
\label{solutions}
\begin{gather}
d_{x1} = -\frac{1}{m_{1}}m_{2}d_{x2}-\frac{1}{m_{1}}m_{3}d_{x3}-\frac{1}{m_{1}} \sum_{n=4}^{N} m_{n} d_{xn} = 0, \\
d_{y1} = -\frac{1}{m_{1}}m_{2}d_{y2}-\frac{1}{m_{1}} \sum_{n=3}^{N} m_{n} d_{yn} = 0, \\
d_{z1} = -\frac{1}{m_{1}} \sum_{n=2}^{N} m_{n} d_{zn} = 0, \\
d_{x2} = -\frac{v^{x}_{z2}}{v^{x}_{x2}} d_{z2} -\frac{v^{x}_{x3}}{v^{x}_{x2}} d_{x3}-\frac{v^{x}_{z3}}{v^{x}_{x2}} d_{z3}
-\frac{1}{v^{x}_{x2}}\sum_{n=4}^{N} \left( v^{x}_{xn}d_{xn}+v^{x}_{zn}d_{zn}\right), \\ 
d_{y2} = -\frac{v^{y}_{z2}}{v^{y}_{y2}}d_{z2} -\frac{1}{v^{y}_{y2}} \sum_{n=3}^{N} \left( v^{y}_{yn}d_{yn}+v^{y}_{zn}d_{zn}
\right), \\
d_{z2} = d_{z2} \\
d_{x3} = -\frac{v^{z}_{y3}}{v^{z}_{x3}}d_{y3}-\frac{v^{z}_{z3}}{v^{z}_{x3}}d_{z3}
 -\frac{1}{v^{z}_{x3}} \sum_{n=4}^{N} \left( v^{z}_{xn}d_{xn}+v^{z}_{yn}d_{yn}+v^{z}_{zn}d_{zn} \right), \\
d_{y3} = d_{y3}, \\
d_{z3} = d_{z3}, \\  
d_{x4} = d_{x4}, \\
d_{y4} = d_{y4}, \\
d_{z4} = d_{z4}, \\
d_{x5} = d_{x5}, \\
\vdots \\
d_{xN} = d_{xN}, \\
d_{yN} = d_{yN}, \\
d_{zN} = d_{zN}.
\end{gather}
\end{subequations}
}
\end{widetext}

With the dependent variables removed from the right-hand sides in Eq.\, (\ref{solutions}) one has
\begin{widetext}
\begin{subequations}
\label{sol2}
\begin{gather}
d_{x1} = \frac{m_{2}}{m_{1}}\frac{v^{x}_{z2}}{v^{x}_{x2}} d_{z2}+ \frac{u_{1}}{m_{1}}\frac{v^{z}_{y3}}{v^{z}_{x3}}d_{y3}
+\left( \frac{u_{1}}{m_{1}} \frac{v^{z}_{z3}}{v^{z}_{x3}}+\frac{m_{2}}{m_{1}}\frac{v^{x}_{z3}}{v^{x}_{x2}}\right) d_{z3} \nonumber \\
+\sum_{n=4} \left[ \left( \frac{u_{1}}{m_{1}} \frac{v_{xn}^{z}}{v_{x3}^{z}} + \frac{m_{2}}{m_{1}} \frac{v_{xn}^{x}}{v_{x2}^{x}}
-\frac{m_{n}}{m_{1}} \right) d_{xn} + \frac{u_{1}}{m_{1}}\frac{v_{yn}^{z}}{v_{x3}^{z}} d_{yn}
+\left( \frac{u_{1}}{m_{1}}\frac{v_{zn}^{z}}{v_{x3}^{z}} +\frac{m_{2}}{m_{1}}\frac{v_{zn}^{x}}{v_{x2}^{x}} \right) d_{zn} \right], \\
\textrm{where} \nonumber \\
u_{1} = m_{3}-m_{2} \frac{v^{x}_{x3}}{v^{x}_{x2}}, \nonumber \\
d_{y1} = \frac{m_{2}}{m_{1}}\frac{v^{y}_{z2}}{v^{y}_{y2}} d_{z2}
+ \sum_{n=3}^{N} \left[ \left( \frac{m_{2}}{m_{1}}\frac{v^{y}_{yn}}{v^{y}_{y2}}-\frac{m_{n}}{m_{1}}\right) d_{yn}
+\frac{m_{2}}{m_{1}}\frac{v^{y}_{zn}}{v^{y}_{y2}} d_{zn} \right], \\
d_{z1} = -\sum_{n=2}^{N} \frac{m_{n}}{m_{1}}d_{zn},  \\
d_{x2} = -\frac{v^{x}_{z2}}{v^{x}_{x2}} d_{z2}+\frac{v^{x}_{x3}}{v^{x}_{x2}}\frac{v^{z}_{y3}}{v^{z}_{x3}}d_{y3}
+\left( \frac{v^{x}_{x3}}{v^{x}_{x2}}\frac{v^{z}_{z3}}{v^{z}_{x3}}-\frac{v^{x}_{z3}}{v^{x}_{x2}}\right) d_{z3} \nonumber \\
+\sum_{n=4}^{N} \left[ \left( \frac{v^{x}_{x3}}{v^{x}_{x2}}\frac{v^{z}_{xn}}{v^{z}_{x3}}-\frac{v^{x}_{xn}}{v^{x}_{x2}}\right)d_{xn}
+\frac{v^{x}_{x3}}{v^{x}_{x2}}\frac{v^{z}_{yn}}{v^{z}_{x3}}d_{yn}
+\left( \frac{v^{x}_{x3}}{v^{x}_{x2}}\frac{v^{z}_{zn}}{v^{z}_{x3}}-\frac{v^{x}_{zn}}{v^{x}_{x2}}\right)d_{zn}
\right], \\
d_{y2} = -\frac{v^{y}_{z2}}{v^{y}_{y2}} d_{z2}-\sum_{n=3}^{N}\left( \frac{v^{y}_{yn}}{v^{y}_{y2}} d_{yn}
+\frac{v^{y}_{zn}}{v^{y}_{y2}} d_{zn} \right), \\
d_{z2} = d_{z2}, \\ 
d_{x3} = -\frac{v^{z}_{y3}}{v^{z}_{x3}} d_{y3}-\frac{v^{z}_{z3}}{v^{z}_{x3}} d_{z3}
-\sum_{n=4}^{N} \left( \frac{v^{z}_{xn}}{v^{z}_{x3}} d_{xn}+\frac{v^{z}_{yn}}{v^{z}_{x3}} d_{yn}
+\frac{v^{z}_{zn}}{v^{z}_{x3}} d_{zn} \right), \\
d_{y3} = d_{y3}, \\
d_{z3} = d_{z3}, \\
d_{x4} = d_{x4}, \\
d_{y4} = d_{y4}, \\
d_{z4} = d_{z4}, \\
d_{x5} = d_{x5}, \\
\vdots \\
d_{xN} = d_{xN}, \\
d_{yN} = d_{yN}, \\
d_{zN} = d_{zN}.
\end{gather}
\end{subequations}
\end{widetext}

By collecting the coefficients of the independent variables 
\begin{widetext}
\[ d_{z2},d_{y3},d_{z3},d_{x4},d_{y4},d_{z4},\ldots,d_{xn},d_{yn},d_{zn},\ldots,d_{xN},d_{yN},d_{zN} \] 
\end{widetext}
given in Eq.\, (\ref{sol2}) into columns one obtains the vectors
\begin{widetext}
\[ \bm{h}^{\textrm{E},1},\bm{h}^{\textrm{E},2},\bm{h}^{\textrm{E},3}, \bm{h}^{\textrm{E},4},\bm{h}^{\textrm{E},5},
\bm{h}^{\textrm{E},6},\ldots,\bm{h}^{\textrm{E},j},\ldots,
\bm{h}^{\textrm{E},3N-8},\bm{h}^{\textrm{E},3N-7},\bm{h}^{\textrm{E},3N-6},
\]
\end{widetext}
respectively, given in Tables\, \ref{hvecs-1} and \ref{hvecs-1-cont}.
One can check by simple analytical calculations employing Eqs.\, (\ref{vx},\ref{vy}) and (\ref{vz}) 
that these $\bm{h}^{\textrm{E},j}$
vectors do, indeed, obey the Eckart conditions Eq.\, (\ref{eckartequations}). 
For instance, by replacing the components of $\bm{d}$ with those of $\bm{h}^{\textrm{E},3}$ in Eq.\, (\ref{rz}) one
obtains that
\begin{gather}
\sum_{n=3}^{N} \left( a^{0}_{xn}m_{n}h^{\textrm{E},3}_{yn}-m_{n}h^{\textrm{E},3}_{xn}a^{0}_{yn}  \right) \nonumber \\ 
=
m_{2} \frac{v^{y}_{z3}}{v^{y}_{y2}} \left( a^{0}_{x1}-a^{0}_{x2}\right) + 
m_{2} \frac{v^{x}_{z3}}{v^{x}_{x2}} \left( a^{0}_{y2}-a^{0}_{y1}\right) \nonumber \\
+ \left[ m_{3} \left( a^{0}_{y3}-a^{0}_{y1}\right)
+m_{2} \frac{v^{x}_{x3}}{v^{x}_{x2}} \left( a^{0}_{y1}-a^{0}_{y2}\right) \right] \frac{v^{z}_{z3}}{v^{z}_{x3}} \nonumber \\
= v^{z}_{z3} - v^{z}_{x3}\frac{v^{z}_{z3}}{v^{z}_{x3}} = 0.
\end{gather}

One may have noticed that some terms can become singular. The singularities can be avoided simply
by reorienting the coordinate system.     

The vectors $\bm{h}^{\textrm{E},j}$ may not be orthonormal with mass weighting, i.e.
$\left[ \bm{h}^{\textrm{E},i}\right]^{T}\bm{m}\bm{h}^{\textrm{E},j}\neq \delta_{ij}$.
One may use the Gram-Schmidt procedure to
orthogonalize the vectors $\bm{m}^{1/2}\bm{h}^{\textrm{E},j}$ to obtain a set of orthonormal vectors $\bm{e}_{j}$:
\begin{subequations}
\label{gram-schmidt}
\begin{gather}
\bm{l}_{1}= \bm{m}^{1/2}\bm{h}^{\textrm{E},1}, ~~~
\bm{e}_{1} = \bm{l}_{1}/\| \bm{l}_{1}\|, \\
\bm{l}_{2}= \left(\bm{I}-\bm{e}_{1}\bm{e}_{1}^{T}\right) \bm{m}^{1/2}\bm{h}^{\textrm{E},2}, ~~~
\bm{e}_{2} = \bm{l}_{2}/ \| \bm{l}_{2}\|, \\
\bm{l}_{3}= \left( \bm{I}-\bm{e}_{1}\bm{e}_{1}^{T}-\bm{e}_{2}\bm{e}_{2}^{T} \right) 
\bm{m}^{1/2}\bm{h}^{\textrm{E},3}, ~~~
\bm{e}_{3} = \bm{l}_{3}/\| \bm{l}_{3}\|, \\
\vdots  \nonumber \\
\bm{l}_{3N-6} =\left( \bm{I} - \sum_{j=1}^{3N-7} \bm{e}_{j}\bm{e}_{j}^{T} \right) \bm{m}^{1/2}\bm{h}^{\textrm{E},3N-6}, ~~~
\bm{e}_{3N-6} = \bm{l}_{3N-6}/\| \bm{l}_{3N-6}\|,
\end{gather}
\end{subequations}
where $\| \bm{e}_{j}\| = \left(\left[ \bm{l}_{j}\right]^{T}\bm{l}_{j}\right)^{1/2}$, and $\bm{I}$ is the 
$3N$ by $3N$ unit matrix.
Then, basis vectors, $\bm{d}^{\textrm{E},j}$, spanning the vibrational space and orthonormal with respect to 
mass weighting can be calculated by the equations
\begin{gather}
\bm{d}^{\textrm{E},j} = \bm{m}^{-1/2}\bm{e}_{j}.
\end{gather} 

\clearpage

\begin{turnpage}
\small

\begin{table*}[!htb] 
\bgroup
\def\arraystretch{1}
\begin{center}
\caption{\label{hvecs-1}Analytical expressions of particular solutions, $\bm{h}^{\textrm{E},j}$, 
of the Eckart conditions of an $N$-atom non-linear molecule}
\begin{gather}
\begin{array}{ccccccc}
 & \bm{h}^{\textrm{E},1} & \bm{h}^{\textrm{E},2} & \bm{h}^{\textrm{E},3} & \bm{h}^{\textrm{E},4} & \bm{h}^{\textrm{E},5} &
\bm{h}^{\textrm{E},6} \\
& d_{z2} & d_{y3} & d_{z3} & d_{x4} & d_{y4}  & d_{z4} \\
\begin{array}{c}
x1 \\
y1 \\
z1 \\
x2 \\
y2 \\
z2 \\
x3 \\
y3 \\
z3 \\
x4 \\
y4 \\
z4 \\
x5 \\
\vdots \\
xN \\
yN \\
zN
\end{array}
&
\left( 
\begin{array}{c}
\frac{m_{2}}{m_{1}}\frac{v^{x}_{z2}}{v^{x}_{x2}} \\
\frac{m_{2}}{m_{1}}\frac{v^{y}_{z2}}{v^{y}_{y2}} \\
-\frac{m_{2}}{m_{1}} \\
-\frac{v^{x}_{z2}}{v^{x}_{x2}} \\
-\frac{v^{y}_{z2}}{v^{y}_{y2}} \\
1 \\
0 \\
0 \\
0 \\
0 \\
0 \\
0 \\
0 \\
\vdots \\
0 \\
0 \\
0 
\end{array}
\right) &
\left( 
\begin{array}{c}
\frac{u_{1}}{m_{1}}\frac{v^{z}_{y3}}{v^{z}_{x3}} \\
\frac{m_{2}}{m_{1}}\frac{v^{y}_{y3}}{v^{y}_{y2}}-\frac{m_{3}}{m_{1}} \\
0 \\
\frac{v^{x}_{x3}}{v^{x}_{x2}}\frac{v^{z}_{y3}}{v^{z}_{x3}} \\
-\frac{v^{y}_{y3}}{v^{y}_{y2}} \\
0 \\
-\frac{v^{z}_{y3}}{v^{z}_{x3}} \\
1 \\
0 \\
0 \\
0 \\
0 \\
0 \\
\vdots \\
0 \\
0 \\
0 
\end{array}
\right)  &
\left(
\begin{array}{c}
\frac{u_{1}}{m_{1}} \frac{v^{z}_{z3}}{v^{z}_{x3}}+\frac{m_{2}}{m_{1}}\frac{v^{x}_{z3}}{v^{x}_{x2}} \\
\frac{m_{2}}{m_{1}}\frac{v^{y}_{z3}}{v^{y}_{y2}} \\
-\frac{m_{3}}{m_{1}} \\
\frac{v^{x}_{x3}}{v^{x}_{x2}}\frac{v^{z}_{z3}}{v^{z}_{x3}}-\frac{v^{x}_{z3}}{v^{x}_{x2}} \\
-\frac{v^{y}_{z3}}{v^{y}_{y2}} \\
0 \\
-\frac{v^{z}_{z3}}{v^{z}_{x3}} \\
0 \\
1 \\
0 \\
0 \\
0 \\
0 \\
\vdots \\
0 \\
0 \\
0 
\end{array}
\right) & 
\left(
\begin{array}{c}
\frac{u_{1}}{m_{1}} \frac{v_{x4}^{z}}{v_{x3}^{z}} + \frac{m_{2}}{m_{1}} \frac{v_{x4}^{x}}{v_{x2}^{x}}
-\frac{m_{4}}{m_{1}} \\
0 \\
0 \\
\frac{v^{x}_{x3}}{v^{x}_{x2}}\frac{v^{z}_{x4}}{v^{z}_{x3}}-\frac{v^{x}_{x4}}{v^{x}_{x2}} \\
0 \\
0 \\
-\frac{v^{z}_{x4}}{v^{z}_{x3}} \\
0 \\
0 \\
1 \\
0 \\
0 \\
0 \\
\vdots \\
0 \\
0 \\
0 
\end{array}
\right) &
\left(
\begin{array}{c}
\frac{u_{1}}{m_{1}}\frac{v_{y4}^{z}}{v_{x3}^{z}}\\
\frac{m_{2}}{m_{1}}\frac{v^{y}_{y4}}{v^{y}_{y2}}-\frac{m_{4}}{m_{1}} \\
0 \\
\frac{v^{x}_{x3}}{v^{x}_{x2}}\frac{v^{z}_{y4}}{v^{z}_{x3}} \\
-\frac{v^{y}_{y4}}{v^{y}_{y2}} \\
0 \\
-\frac{v^{z}_{y4}}{v^{z}_{x3}} \\
0 \\
0 \\
0 \\
1 \\
0 \\
0 \\
\vdots \\
0 \\
0 \\
0
\end{array}
\right) &
\left( 
\begin{array}{c}
\frac{u_{1}}{m_{1}}\frac{v_{z4}^{z}}{v_{x3}^{z}} +\frac{m_{2}}{m_{1}}\frac{v_{z4}^{x}}{v_{x2}^{x}}\\
\frac{m_{2}}{m_{1}}\frac{v^{y}_{z4}}{v^{y}_{y2}} \\
-\frac{m_{4}}{m_{1}} \\
\frac{v^{x}_{x3}}{v^{x}_{x2}}\frac{v^{z}_{z4}}{v^{z}_{x3}}-\frac{v^{x}_{z4}}{v^{x}_{x2}} \\
-\frac{v^{y}_{z4}}{v^{y}_{y2}} \\
0 \\
-\frac{v^{z}_{z4}}{v^{z}_{x3}} \\
0 \\
0 \\
0 \\
0 \\
1 \\
0 \\
\vdots \\
0 \\
0 \\
0
\end{array} 
\right)
\end{array}
\nonumber
\end{gather}
\end{center}
\egroup
\end{table*}

\end{turnpage}

\clearpage

\begin{turnpage}
\small
\begin{table*}[!htb]
\bgroup
\def\arraystretch{1}
\begin{center}
\caption{\label{hvecs-1-cont} Table\, \ref{hvecs-1} continued}
\begin{gather}
\begin{array}{cccccc}
& \bm{h}^{\textrm{E},7} & \bm{h}^{\textrm{E},j} & \bm{h}^{\textrm{E},3N-8} & \bm{h}^{\textrm{E},3N-7} & \bm{h}^{\textrm{E},3N-6} \\
& d_{x5 }               &                       & d_{xN}                   & d_{yN}                   & d_{zN} \\
\begin{array}{c}
x1 \\
y1 \\
z1 \\
x2 \\
y2 \\
z2 \\
x3 \\
y3 \\
z3 \\
x4 \\
y4 \\
z4 \\
x5 \\
\vdots \\
xN \\
yN \\
zN 
\end{array}
&
\left(
\begin{array}{c}
\frac{u_{1}}{m_{1}} \frac{v_{x5}^{z}}{v_{x3}^{z}} + \frac{m_{2}}{m_{1}} \frac{v_{x5}^{x}}{v_{x2}^{x}}
-\frac{m_{5}}{m_{1}} \\
0 \\
0 \\
\frac{v^{x}_{x3}}{v^{x}_{x2}}\frac{v^{z}_{x5}}{v^{z}_{x3}}-\frac{v^{x}_{x5}}{v^{x}_{x2}} \\
0 \\
0 \\
-\frac{v^{z}_{x5}}{v^{z}_{x3}} \\
0 \\
0 \\
0 \\
0 \\
0 \\
1 \\
\vdots \\
0 \\
0 \\
0 
\end{array}
\right) 
& 
\left(
\begin{array}{c}
\cdot \\
\cdot \\
\cdot \\
\cdot \\
\cdot \\
\cdot \\
\cdot \\
\cdot \\
\cdot \\
\cdot \\
\cdot \\
\cdot \\
\cdot \\
\vdots \\
\cdot \\
\cdot \\
\cdot
\end{array}
\right) 
& 
\left(
\begin{array}{c}
\frac{u_{1}}{m_{1}} \frac{v_{xN}^{z}}{v_{x3}^{z}} + \frac{m_{2}}{m_{1}} \frac{v_{xN}^{x}}{v_{x2}^{x}}
-\frac{m_{N}}{m_{1}} \\
0 \\
0 \\
\frac{v^{x}_{x3}}{v^{x}_{x2}}\frac{v^{z}_{xN}}{v^{z}_{x3}}-\frac{v^{x}_{xN}}{v^{x}_{x2}} \\
0 \\
0 \\
-\frac{v^{z}_{xN}}{v^{z}_{x3}} \\
0 \\
0 \\
0 \\
0 \\
0 \\
0 \\
\vdots \\
1  \\
0 \\
0
\end{array}
\right) &
\left(                          
\begin{array}{c}
\frac{u_{1}}{m_{1}}\frac{v_{yN}^{z}}{v_{x3}^{z}}\\
\frac{m_{2}}{m_{1}}\frac{v^{y}_{yN}}{v^{y}_{y2}}-\frac{m_{N}}{m_{1}} \\
0 \\
\frac{v^{x}_{x3}}{v^{x}_{x2}}\frac{v^{z}_{yN}}{v^{z}_{x3}} \\
-\frac{v^{y}_{yN}}{v^{y}_{y2}} \\
0 \\
-\frac{v^{z}_{yN}}{v^{z}_{x3}} \\
0 \\
0 \\
0 \\
0 \\
0 \\
0 \\
\vdots \\
0 \\
1 \\
0 
\end{array}
\right) &
\left(
\begin{array}{c}
\frac{u_{1}}{m_{1}}\frac{v_{zN}^{z}}{v_{x3}^{z}} +\frac{m_{2}}{m_{1}}\frac{v_{zN}^{x}}{v_{x2}^{x}}\\
\frac{m_{2}}{m_{1}}\frac{v^{y}_{zN}}{v^{y}_{y2}} \\
-\frac{m_{N}}{m_{1}} \\
\frac{v^{x}_{x3}}{v^{x}_{x2}}\frac{v^{z}_{zN}}{v^{z}_{x3}}-\frac{v^{x}_{zN}}{v^{x}_{x2}} \\
-\frac{v^{y}_{zN}}{v^{y}_{y2}} \\
0 \\
-\frac{v^{z}_{zN}}{v^{z}_{x3}} \\
0 \\
0 \\
0 \\
0 \\
0 \\
0 \\
\vdots \\
0 \\
0 \\
1
\end{array}
\right)
\end{array}
\nonumber
\end{gather}
\end{center}
\egroup
\end{table*}

\end{turnpage}

\section{Optimal Eckart displacements} 
\label{iterativemin}

For minimizing 
$\textrm{MWSD}$ of Eq.\, (\ref{form1}) a neater expression than Eq.\, (\ref{form1}) arises by considering that
\begin{gather}
c_{j} = \sum_{n=1}^{N} m_{n} \bm{d}^{\textrm{E},j}_{n} \cdot
\left( \bm{U} \bm{a}_{n}-\bm{a}_{n}^{0}\right).
\end{gather} 
Since a dot product $\bm{u}\cdot \bm{v}$ 
can be written as 
\begin{widetext}
\begin{gather}
2\bm{u}\cdot \bm{v} = \lVert \bm{u} + \bm{v} \rVert^{2}
- \lVert \bm{u} \rVert^{2} -\lVert \bm{v} \rVert^{2},
\end{gather}
\end{widetext}
 one obtains
 \begin{widetext}
 \begin{gather}
 2c_{j} = \sum_{n=1}^{N} m_{n}
 \lVert  \bm{d}^{\textrm{E},j}_{n} + \bm{U} \bm{a}_{n}-\bm{a}_{n}^{0}  
 \rVert^{2} 
 -1
 - \sum_{n=1}^{N} m_{n} 
 \lVert \bm{U}\bm{a}_{n}-\bm{a}_{n}^{0} \rVert^{2},
 \label{cdef}
 \end{gather}
 \end{widetext}
 where use has been made of the equality
 \begin{gather}
 \sum_{n=1}^{N} m_{n} 
  \lVert \bm{d}^{\textrm{E},j}_{n} \rVert^{2} = 1.
 \end{gather}
 
To carry on recall that a rotation matrix $\bm{U}$ may be parametrized in terms
of the scalar, $q_{0}$, and vector, 
$\bm{q} =\left( q_{1}, q_{2},q_{3} \right) $, components
of a quaternion $Q = \left[q_{0}, \bm{q} \right] $ of unit norm, and a
rotated vector $\bm{U}\bm{a}_{n}$ can be calculated as products of quaternions, since
\begin{gather}
\left[ 0, \bm{U}\bm{a}_{n} \right] = Q^{-1}A_{n}Q,
\end{gather}
where $\left[ 0, \bm{U}\bm{a}_{n} \right]$ is the pure quaternion corresponding 
to the vector $\bm{U}\bm{a}_{n}$, and $A_{n} = \left[0,\bm{a}_{n} \right]$ denotes the pure
quaternion corresponding to the vector $\bm{a}_{n}$. 

Therefore we can rewrite Eq.\, (\ref{cdef})
as
\begin{widetext}
\begin{gather}
 2c_{j} = \sum_{n=1}^{N} m_{n}
 \lVert  D^{\textrm{E},j}_{n} + Q^{-1}A_{n}Q-A_{n}^{0}   
 \rVert^{2} 
 -\lVert Q^{-1}Q\rVert^{2}
 - \sum_{n=1}^{N} m_{n} 
 \lVert Q^{-1}A_{n}Q-A_{n}^{0} \rVert^{2},
 \nonumber \\
 = \sum_{n=1}^{N} m_{n}
  \lVert   A_{n}Q-Q\left( A_{n}^{0}- D^{\textrm{E},j}_{n}\right)  
  \rVert^{2} 
  -\lVert Q\rVert^{2}
  - \sum_{n=1}^{N} m_{n} 
  \lVert A_{n}Q-QA_{n}^{0} \rVert^{2},
\label{cdefquat}
\end{gather}
\end{widetext} 
where $ D^{\textrm{E},j}_{n}$ is the pure quaternion corresponding to the vector $\bm{d}_{n}^{\textrm{E},j}$, 
and
the second equality follows from the fact that the norm of the product of quaternions is 
equal to the product of their norm. By expressing the quaternions in Eq.\, (\ref{cdefquat})
in terms of their components one can see that $c_{j}$ is a
quadratic form
\begin{gather}
c_{j}= \sum_{r,t=0}^{3} C_{rt}^{j}q_{r}q_{t} = \bm{q}^{T}\bm{C}^{j} \bm{q}, \label{cquad}
\end{gather} 
with $\bm{q}^{T}=\left(q_{0},q_{1},q_{2},q_{3}\right)$ 
denoting a row vector containing the components of the quaternion $Q$. 

Therefore,
\begin{gather}
\textrm{MWSD} = \sum_{j=1}^{3N-6} \left[\bm{q}^{T}\bm{C}^{j} \bm{q}\right]^{2} \label{mwsed}
\end{gather} 
and it has to be minimized with respect to $\bm{q}$ under the normalization condition $ \lVert \bm{q} \rVert =1$.
The optimal $\bm{q}$ may be determined iteratively.

Note that
\begin{gather}
\bm{q} = \left( 
\begin{array}{c}
\cos \theta \\
n_{1} \sin \theta \\
n_{2} \sin \theta\\
n_{3} \sin \theta
\end{array}
\right) =
\left( 
\begin{array}{cccc}
1 & 0     & 0     & 0  \\
0 & n_{1} & 0     & 0  \\    
0 & 0     & n_{2} & 0 \\
0 & 0     & 0     & n_{3}
\end{array}
\right)
\left(
\begin{array}{c}
 \cos \theta \\
 \sin \theta \\
 \sin \theta \\
 \sin \theta 
\end{array}
\right).
\end{gather}
Therefore,
\begin{widetext}
\begin{gather}
\textrm{MWSD}=\sum_{j}\left\{ C^{(j)}_{00}\cos^{2}\theta + \left[ \sum_{i=1}^{3} \left( 
C^{(j)}_{i0}+C^{(j)}_{0i}\right) n_{i}\right] 
\sin\theta\cos\theta + \left[ \sum_{i,k=1}^{3}
n_{i}C^{(j)}_{ik}n_{k} \right] \sin^{2}\theta \right\}^{2},
\end{gather}
\end{widetext}
and we can find $\theta$ which minimizes MWSD at fixed $\bm{n}$.
Next update $\theta$ and find the optimal $\bm{n}$ at this new, fixed $\theta$.  

Note that
\begin{widetext}
\begin{gather}
\bm{n} = \left(
\begin{array}{c}
n_{1} \\
n_{2} \\
n_{3}
\end{array}
\right) =
\left(
\begin{array}{c}
\cos\chi \sin\phi \\
\sin\chi \sin\phi \\
\cos\phi
\end{array}
\right) = 
\left(
\begin{array}{ccc}
\sin\phi & 0 & 0 \\
0 &   \sin\phi & 0 \\
0 &  0  & \cos\phi 
\end{array}
\right)
\left(
\begin{array}{c}
\cos\chi \\
\sin\chi \\
1
\end{array}
\right)
\end{gather}
\end{widetext}

Thus, with $\theta$ and $\phi$ fixed, we can find the optimal $\chi$ by 
minimizing
\begin{widetext} 
\begin{gather}
\textrm{MWSD}= \sum_{j}\left\{
\left[C^{(j)}_{00}\cos^{2}\theta +
\left( C^{(j)}_{03}+
C^{(j)}_{30}\right) \sin\theta\cos\theta \cos\phi
+C^{(j)}_{33}\sin^{2}\theta\cos^{2}\phi \right] \right. \nonumber \\ \left. 
+\left[
\left( C^{(j)}_{01}+C^{(j)}_{10}\right) 
\sin\theta\cos\theta \sin\phi 
+ \left( C^{(j)}_{13} + C^{(j)}_{31} \right)\sin^{2}\theta\sin\phi\cos\phi 
\right]\cos\chi \right.\nonumber \\ \left.
+\left[
\left( C^{(j)}_{02}+C^{(j)}_{20}\right) 
\sin\theta\cos\theta \sin\phi 
+\left(  C^{(j)}_{23} +C^{(j)}_{32} \right)\sin^{2}\theta \sin\phi\cos\phi
\right]\sin\chi \right. \nonumber \\ \left.
+\left[ 
\left( C^{(j)}_{12} +
C^{(j)}_{21} \right) \sin^{2}\theta\sin^{2}\phi \right] \sin\chi\cos\chi \right. \nonumber \\ \left.
+
\left[ 
C^{(j)}_{11}\sin^{2}\theta\sin^{2}\phi \right]\cos^{2}\chi
\right. \nonumber \\ \left. 
+
\left[
C^{(j)}_{22}\sin^{2}\theta\sin^{2}\phi\right] \sin^{2}\chi
\right\}^{2}
\end{gather}
\end{widetext}

Now having updated $\chi$, with $\theta$ and $\chi$ fixed we look for the optimal 
$\phi$.  By using 
\begin{widetext}
\begin{gather}
\bm{n} = \left(
\begin{array}{c}
n_{1} \\
n_{2} \\
n_{3}
\end{array}
\right) =
\left(
\begin{array}{c}
\cos\chi \sin\phi \\
\sin\chi \sin\phi \\
\cos\phi
\end{array}
\right) = 
\left(
\begin{array}{ccc}
\cos\chi & 0 & 0 \\
0 &   \sin\chi & 0 \\
0 &  0  & 1 
\end{array}
\right)
\left(
\begin{array}{c}
\sin\phi \\
\sin\phi \\
\cos\phi
\end{array}
\right)
\end{gather}
\end{widetext}
it can be derived that 
\begin{gather}
\textrm{MWSD}= \sum_{j} \left\{
C^{(j)}_{00}\cos^{2}\theta \right. \nonumber \\ \left.
+\left[
\left(C^{(j)}_{03}+ C^{(j)}_{30} \right) \sin\theta\cos\theta +
\left(C^{(j)}_{01}+ C^{(j)}_{10}\right) \sin\theta\cos\theta \cos\chi \right] \cos\phi \right. \nonumber \\
\left. +\left[ \left( C^{(j)}_{02}+C^{(j)}_{20} \right) 
\sin\theta\cos\theta \sin\chi
\right] \sin\phi \right. \nonumber \\ \left.
+
\left[ 
\left( C^{(j)}_{13}+C^{(j)}_{31} \right) \sin^{2}\theta \cos\chi
+\left( C^{(j)}_{23} + C^{(j)}_{32} \right)\sin^{2}\theta\sin\chi \right] \sin\phi\cos\phi
\right. \nonumber \\ \left.
+\left[
C^{(j)}_{33}\sin^{2}\theta \right]\cos^{2}\phi \right. \nonumber \\ \left.
+\left[
\left( C^{(j)}_{11} +C^{(j)}_{22}\right)\sin^{2}\theta \sin^{2}\chi +
\left( C^{(j)}_{12}+C^{(j)}_{21}\right)\sin^{2}\theta \sin\chi\cos\chi \right] \sin^{2}\phi
\right\}.
\end{gather}
Minimization of this function of $\phi$ gives the optimal $\phi$. Then, update $\phi$ and iterate the procedure until convergence.

The 1D optimizations required are reduced to finding zeros of functions of single variable. 
To carry out calculations a MAPLE\cite{maple} code has been
written.

\clearpage

\begin{thebibliography}{99}

\bibitem{sutcliffe-hinchliffe} B. T. Sutcliffe,
Calculations of the Vibration-Rotation Spectra of Small Molecules, in 
Specialist Periodical Reports, Chemical Modelling, Applications and Theory, Edited by A. Hinchliffe,
 (The Royal Society of Chemistry, 2004), Vol. 3, pp. 1-44.
\bibitem{eckart} C. Eckart, Phys. Rev. {\bf 47}, 552 (1935).
\bibitem{watson} J. K. G. Watson, Mol. Phys. {\bf 15}, 479 (1968).


\bibitem{meyer-gunthard} R. Meyer and Hs. H. G\"unthard, J. Chem. Phys. {\bf 49}, 1510 (1968).
\bibitem{pickett} H. M. Pickett, J. Chem. Phys. {\bf 56}, 1715 (1972).



\bibitem{wei-carrington-1} H. Wei and T. Carrington, Jr., J. Chem. Phys. {\bf 107}, 2813 (1997).
\bibitem{wei-carrington-2} H. Wei and T. Carrington, Jr., J. Chem. Phys. {\bf 107}, 9493 (1997).
\bibitem{mardis-sibert} K. L. Mardis and E. L. Sibert III, J. Chem. Phys. {\bf 106}, 6618 (1997). 
\bibitem{wei-carrington-3} H. Wei and T. Carrington, Jr., Chem. Phys. Lett. {\bf 287}, 289 (1998).
\bibitem{rey}R. Rey, Chem. Phys. {\bf 229}, 217 (1998).
\bibitem{wei} H. Wei, J. Chem. Phys. {\bf 118}, 7208 (2003).
\bibitem{meremianin-1} A. V. Meremianin, J. Chem. Phys. {\bf 120}, 7861 (2004).
\bibitem{pesonen} J. Pesonen, J. Chem. Phys. {\bf 140}, 074101 (2014).


\bibitem{mccoy-burleigh-sibert} A. B. McCoy, D. C. Burleigh, and E. L. Sibert, J. Chem. Phys. {\bf 95}, 7449 (1991).
\bibitem{wang-sibert} X.-G. Wang and E. L. Sibert, Spectrochim Acta Part A {\bf 58}, 863 (2002).
\bibitem{szidarovszky-fabri-csaszar} T. Szidarovszky, C. F\'abri, and A. G. Cs\'asz\'ar,
J. Chem. Phys. {\bf 136}, 174112 (2012).
\bibitem{wang-carrington} X.-G. Wang and T. Carrington, Jr., J. Chem. Phys. {\bf 138}, 104106 (2013).
\bibitem{fabri-matyus-csaszar} C. F\'abri, E. M\'atyus, and A. G. Cs\'asz\'ar, 
Spectrochim Acta Part A {\bf 119}, 84 (2014).
\bibitem{sadri} K. Sadri, D. Lauvergnat, F. Gatti, and H.-D. Meyer, 
J. Chem. Phys. {\bf 141}, 114101 (2014).
\bibitem{yachmenev-yurchenko} A. Yachmenev and S. N. Yurchenko
J. Chem. Phys. {\bf 143}, 014105 (2015).
\bibitem{lauvergnat-luis} D. Lauvergnat, J. M. luis, B. Kirtman, H. Reis, and A. Nauts,
J. Chem. Phys. {\bf 144}, 084116 (2016).


\bibitem{h5+} T. C. Cheng, L. Jiang, K. R. Asmis, Y Wang, J. M. Bowman, A. M. Ricks, and M. A. Duncan, 
J. Phys. Chem. Lett.  {\bf 3}, 3160 (2012).
\bibitem{ch5+} O. Asvany, K. M. T. Yamada, S. Brünken, A. Potapov, and
S. Schlemmer, Science {\bf 347}, 1346 (2015).
\bibitem{h2o}  P. Ayotte, J. A. Kelley, S. B. Nielsen, and M. A. Johnson, Chem.
Phys. Lett., {\bf 316}, 455 (2000).
\bibitem{ch4h2o-1} R. D. Suenram, G. T. Fraser, F. J. Lovas, and Y. Kawashima,
J. Chem. Phys., 1994, {\bf 101}, 7230 (1994).
\bibitem{ch4h2o-2} L. Dore, R. C. Cohen, C. A. Schmuttenmaer, K. L. Busarow,
M. J. Elrod, J. G. Loeser, and R. J. Saykally, J. Chem. Phys., {\bf 100}, 863 (1994).



\bibitem{sueur} C. R. le Sueur, S. Miller, J. Tennyson, and B. T. Sutcliffe, Mol. Phys. {\bf 76},
1147 (1992).
\bibitem{noguti-gon} T. Noguti and N. Go, J. of Phys. Chem. Soc. of Japan, {\bf 52}, 3283 (1983).


\bibitem{changala} NITROGEN, Numerical and Iterative Techniques for Rovibronic Energies 
with General Internal Coordinates, a program by P. B. Changala, http://www.colorado.edu/nitrogen.
\bibitem{wang-song} L.-P. Wang and C. Song, J. Chem. Phys. {\bf 144}, 214108 (2016).
\bibitem{changala-baraban} P. B. Changala and J. H. Baraban, J. Chem. Phys. {\bf 145}, 174106 (2016).


\bibitem{strauss} H. M. Pickett and H. L. Strauss, J. Am. Chem. Soc. {\bf 92}, 7281 (1970).
\bibitem{redding} R. W. Redding and J. T. Hougen, J. Mol. Spectrosc. {\bf 37}, 366 (1971).
\bibitem{meyer} F. O. Meyer and R.W. Redding, J. Mol. Spectrosc. {\bf 70}, 410 (1978).
\bibitem{redding2} R. W. Redding and F. O. Meyer III, J. Mol. Spectrosc. {\bf 74}, 486 (1979).
\bibitem{kudin} K. N. Kudin and A. Y. Dymarsky, J. Chem. Phys. {\bf 122}, 224105 (2005).
\bibitem{dierksen} M. Dierksen, J. Chem. Phys. {\bf 122}, 227101 (2005).
\bibitem{dymarsky} A. Y. Dymarsky and K. N. Kudin, J. Chem. Phys. {\bf 122}, 227102 (2005)
\bibitem{galbraith} J. D. Louck and H. W. Galbraith, Rev. Mod. Phys. {\bf 48}, 69 (1976).
\bibitem{krasnos} S. V. Krasnoshchekov, E. V. Isayeva, and N. F. Stepanov, J. Chem. Phys. {\bf 140}, 154104 (2014).

\bibitem{szalay-eckart} V. Szalay, J. Chem. Phys. {\bf 140}, 234107 (2014).
\bibitem{szalay-gateway} V. Szalay, J. Chem. Phys. {\bf 142}, 174107 (2015).
\bibitem{szalay-aspects} V. Szalay, J. Chem. Phys. {\bf 143}, 064104 (2015).

\bibitem{carl-meyer} C. D. Meyer, Matrix Analysis and Applied Linear Algebra (SIAM, 2001).


\bibitem{wilson} E. B. Wilson, P. C. Cross, and J. C. Decius, 
Molecular Vibration: The Theory of Infrared and Raman Vibrational Spectra ( New York: Dover, 1980).


\bibitem{thompson} H. B. Thompson, J. Chem. Phys. {\bf 47}, 3407 (1967).
\bibitem{hilderbrandt} R. L. Hilderbrandt, J. Chem. Phys. {\bf 51}, 1654 (1969).  
\bibitem{lopata-kiss} A. Lopata and \'A. I. Kiss, Computers Chem.  {\bf 3}, 107 (1979). 
\bibitem{trigub} L. P. Trigub and Yu. A. Kruglyak, J. Struct. Chem. {\bf 24}, 161 (1983).
\bibitem{essen} H. Ess\'en and M. Svensson,  Computers Chem.  {\bf 20}, 389 (1996).

\bibitem{littlejohn-reinsch} R. G. Littlejohn and M. Reinsch, Rev. Mod. Phys. {\bf 69}, 213 (1997).

\bibitem{colbert-miller} D. T. Colbert and W. H. Miller, J. Chem. Phys. {\bf 96}, 1982 (1992).
\bibitem{szalay-smith} V. Szalay and S. C. Smith, J. Chem. Phys. {\bf 110}, 72 (1999).
\bibitem{littlejohn-cargo} R. G. Littlejohn and M. Cargo, J. Chem. Phys. {\bf 116}, 7350 (2002). 
\bibitem{jerke} J. L. Jerke, Y. Lee, and C. J. Tymczak, J. Chem. Phys. {\bf 143}, 064108 (2015).

\bibitem{lanczos} C. Lanczos,J. Res. Nat’l Bur. Std. {\bf 45}, 255 (1950).
\bibitem{cullum} K. Cullum and R. A. Willoughby, 
Lanczos Algorithms for Large Symmetric Eigenvalue Computations (Birkhäuser, Boston, 1985), Vols. 1 and 2.

\bibitem{bramley-carrington} M. J. Bramley and T. Carrington Jr., J. Chem. Phys., {\bf 99}, 8519 (1993).


\bibitem{jensen} P. Jensen, J. Mol. Spectrosc. {\bf 133}, 438 (1989).

\bibitem{johnson-reinhardt} B. R. Johnson and W. P. Reinhardt, J. Chem. Phys. {\bf 85}, 4538 (1986).


\bibitem{fernley-et-al} J. A. Fernley, S. Miller, and J. Tennyson, J. Mol. Spectrosc. {\bf 150}, 597 (1991).


\bibitem{szekeres} P. Szekeres, A Course in Modern Mathematical Physics (Cambridge University Press, 2004). 
\bibitem{schutz} B. Schutz, Geometrical methods of mathematical physics (Cambridge University Press, 1980).

\bibitem{duschinsky} F. Duschinsky, Acta Physicochimica U.R.S.S. {\bf 7}, 551 (1937). 
\bibitem{meier} P. Meier, D. Oschetzki, R. Berger, and G. Rauhut, J. Chem. Phys. {\bf 140}, 18411 (2014).


\bibitem{maple} MAPLE 12, Maple is a registered trademark of Waterloo Maple Software.

\end{thebibliography}
\end{document}